\documentclass[useAMS, usenatbib, usegraphicx]{mn2e}
\usepackage{amssymb}

\def\alf{Alfv\'en\,}
\def\bq{\begin{equation}}
\def\eq{\end{equation}}
\def\ee #1 {\times 10^{#1}}
\def\ut #1 #2 { \, \rmn{#1}^{#2}}
\def\u #1 { \, \rmn{#1}}

\let\grad=\nabla
\newcommand\cross{\bmath{\times}}

\newcommand\sigmaperp{\sigma_\perp}
\newcommand\mri{magnetorotational instability\,}
\def\curl{{\grad \cross}}
\def\div #1 {\grad \cdot #1}

\def\ds{\bmath{ds}}
\def\A{\bmath{A}}

\def\b{\bmath{b}}
\def\e{\bmath{e}}
\def\v{\bmath{v}}

\def\vi{\bmath{v}_i}
\def\vj{\bmath{v}_j}
\def\ve{\bmath{v}_e}
\def\vi{\bmath{v}_i}

\def\vn{\bmath{v}_n}
\def\vk{\bmath{v}_K}
\def\VB{\bmath{V}_B}
\def\B{\bmath{B}}
\def\E{\bmath{E}}            
\def\Fj{\bmath{F}_j}            
\def\FH{\bmath{F}_H}            
\def\Bh{\bmath{\hat{B}}}
\def\fh{\bmath{\hat{\phi}}}  
\def\zh{\bmath{\hat{z}}}     

\def\vk{v_{K}}               

\def\Ep{\bmath{E'}}  
\def\Epa{\bmath{E'_\parallel}}  
\def\Epe{\bmath{E'_\perp}}  
\def\J{\bmath{J}}

\def\dB{\bmath{\delta\B}}

\def\dvp{\bmath{\delta\v_{\perp}}}

\def\dEp{\bmath{\delta\E'_{\perp}}}
\def\dBp{\bmath{\delta\B_{\perp}}}
\def\dJp{\bmath{\delta\J_{\perp}}}
\def\sigv{<\sigma v>}

\newcommand{\delt} [1] {\frac{\partial #1}{\partial t}}

\def\Omh{\hat{\Omega}}

\def\b1{{\bar{\omega}}}
\def\bo2{\bar{\omega}^2}

\title{Ion dynamics and the magnetorotational instability in 
weakly-ionized discs}
\author[B.P.Pandey and Mark Wardle]
         {B.P. Pandey and Mark Wardle \\
{Department of Physics, Macquarie University, Sydney, NSW 2109, Australia}
}
\date{\today}
\pagerange{\pageref{firstpage}--\pageref{lastpage}}
\pubyear{2004}
\begin{document}
\maketitle
\label{firstpage}
\begin{abstract}
The magnetorotational instability (MRI) of a weakly ionized,
differentially rotating, magnetized plasma disc is investigated in the
multi-fluid framework.  The disc is threaded by a uniform vertical
magnetic field and charge is carried by electrons and ions only.  The
inclusion of ion inertia causes significant modification to the
conductivity tensor in a weakly ionized disc. The parallel, Pedersen
and Hall component of conductivity tensor becomes time dependent
quantities resulting in ac and dc components of the conductivity. The time dependence 
of the conductivity causes significant modification to the parameter window of \mri.

The effect of ambipolar and Hall diffusion on the linear growth of the
\mri is examined in the presence of time dependent conductivity
tensor.  We find that the growth rate in the ambipolar regime can become
somewhat larger than the rotational frequency, especially when the
departure from ideal MHD is significant.  Further, the instability
operates on large scale lengths.  This has important implication for 
angular momentum transport in the disc.

When charged grains are the dominant ions, their inertia will play important
role near the mid plane of the protoplanetary discs. Ion inertia
could also be important in transporting angular momentum in 
accretion discs around compact objects, in cataclysmic variables. For example, in
cataclysmic variables, where mass flows from a companion main
sequence star on to a white dwarf, the ionization fraction in the disc
can vary in a wide range. The ion inertial effect in such a disc could significantly 
modify the \mri and threfore, this instability could be a possible driver of the 
observed turbulent motion.

\end{abstract}


\section[]{Introduction}

Angular momentum transport has long been recognised as a key issue in
accretion disc theories \citep{LB, SS}.  However, until the 1990s, a
viable physical mechanism necessary to facilitate this transport in
the absence of tidal effects or gravitational instabilities was
unknown.  The Balbus-Hawley (or magnetorotational) instability
\citep{VV, CH} was proposed \citep{bb1, bb2, bb3} as a viable
mechanism that can efficiently drive MHD turbulence and transport
angular momentum in the disc.  This opened the door for its application
to a wide variety of astrophysical discs.  The requirement for the
\mri to operate in such a disc is that the ambient magnetic field is subthermal 
at the disc midplane and is well coupled to
the disc matter. Although, the lower bound on the weak, subthermal field has never been specified, in recent
work this issue has been addressed in the framework of fully ionized, collisionless
cold electron-ion plasma \citep{kzw}. For a highly ionized disc, the requirement of a weak, subthermal field is easily
satisfied and the \mri grows at the rotation frequency $\Omega$ of the
disc as a low frequency \alf mode with $k\,V_A \sim \Omega$, where $k$
is the wavenumber and $V_A$ is the \alf velocity.  However, many
astrophysical discs are not well coupled to the magnetic field.
Circumstellar, Protoplanetary (PPD), Dwarf Novae (DN), and,
proto-neutron-star discs are good examples of weakly ionized discs
with very low to low (PPDs) and high (DNs) fractional ionization. 
In PPDs for example, the sources of ionization are limited
to the disc surface and \mri may operate only in the outer envelope of
the disc \citep{g1} unless some nonthermal source of ionization viz.
the collision of the energetic electrons with neutrals or \mri induced
turbulent convective homogenization of the entire disc \citep{si} is
assumed.  DN discs are thought to have both hot and fully ionized
accretion state as well as cold and mostly neutral accretion states
\citep{CZ, GM}.  Therefore, the direct application of \cite{bb1}
results are difficult in a weakly ionized disc.

The effect of non-ideal MHD on the \mri has been investigated by
several authors: in the ambipolar regime \citep{bb}, hereafter BB94,
\citep{m1, h3, kz}, the resistive regime \citep{j96, pte, bb3, s1, s2,
fsh, s3, s9} and the Hall regime \citep{w4} W99 hereafter;
\citep{b6}, BT01 hereafter; \citep{s4, s5, sw1, sw2, de}.  At the
densities relevant to cloud cores, ambipolar and Hall diffusion plays
an important role in the transport of mass and angular momentum
\citep{w5, b6}.  W99 and BT01 found that collision of neutrals with
the ionized gas in a weakly ionized disc determines the relative
importance of Ambipolar, Hall or Ohmic diffusion on the \mri.  The
ambipolar and Hall effects are particularly important when the
ionization in the disc is very low and the departure from ideal MHD is
severe.

The dynamics of a weakly ionized disc was investigated in the limit of
zero inertia of the ionized plasma components by W99 and BT01.  This
is usually an excellent approximation when the fractional ionization
is low, and allows the ionized components of the fluid (viz electrons,
ions and grains) to be treated on an equal footing.  However, there
are situations -- even in the low fractional ionisation limit -- where
the inertia of the charged species is important in determining their
drift with respect to the neutral component and hence the diffusion of
the magnetic field. In the weakly-ionised limit this becomes important
when the inertial terms in the ion equation of motion start to compete
with the magnetic and neutral collision terms, in other words when the
disc frequency $\Omega$ becomes comparable to or exceeds the
collision frequency with neutrals \emph{and} the gyrofrequency.
This effect may become important for charged dust grains
because their large mass implies low collision and gyrofrequencies.
For example, in dense PPDs, the dominant ion species are positively
charged grains (especially when $ \sim n_H \ge
10^{11}\,\mbox{cm}^{-3}$), \citep{w5}. In the high fractional
ionization regions also, e.g. near the surface of a magnetic
cataclysmic variable, ion inertia may become important \citep{wa95}.


The effect of ion-inertia on the \mri in a two-fluid framework was
considered by BB94.  The magnetic flux was assumed frozen into the
plasma component in their formulation.  This limits the applicabilty
of their results to the ambipolar diffusion regime. We know from W99 and BT01 that 
Hall effect can compete with 
ambipolar diffusion in the weakly ionized regions of the disc.  In the present work 
we adopt a three component model -- neutrals, ions and electrons where by ``ion''
and ``electron'' we mean the most massive and least massive charged
species, whether positively or negatively charged -- to show that ion
inertial effects modify the growth rate substantially and increases
the parameter window in which the instability may operate.  In
particular, the growth rate in the presence of ion inertia may exceed
the Oort A value ($0.75\,\Omega$) limit.  Growth rates larger than the
Oort A limit have recently been reported for collisionless plasmas in
the presence of plasma kinetic/viscous effects \citep{q1, shm, bb4},
where kinetic and MHD effects combine to give a high growth rate
($1.7\,\Omega$) and shift the fastest growing modes towards longer
wavelengths. 

We shall give a general formulation of the problem
that will not only cover the regions of applicability of BB94 and W99
but also cover the unexplored regions.  This paper revisits the
magnetorotational instability in a weakly ionized disc (W99) by
incorporating the effect of ion inertia.  In section 2, we discuss
formulation of BB94 and compare and contrast the region of
applicability of BB94, W99 in the context of present work.  In section
3, a general formulation of a weakly ionized, near-Keplerian,
magnetized disc is given.  In section 4 we describe equilibrium and
linearization of the disc and derive the dispersion
relation.  Section 5 discusses the energetics of a weakly
ionized, magnetized disc.  In this section, we first discuss the role
played by the ambipolar and Hall terms in the neutral dynamics.
Further, using energy arguments, we discuss the conditions under which
ambipolar diffusion can proceed without dissipation and when ambipolar
diffusion can destabilize the disc.  We also discuss how theHall term can
destabilize the \mri. The properties of wave helicity, is derived.
In section 6, we give the detailed numerical solution of the
dispersion relation.  In section 7, application of the result to
various astrophysical discs is discussed.  Section 8 presents a summary
of our results.

\section{Ion inertia and BB94}
%
BB94 adopted the ambipolar diffusion approximation, in which the
magnetic field is frozen into an electrically neutral ionised plasma
coupled by collisions to the neutral fluid.

The effect of ion
inertia on the dynamics ($\rho_i\,d_t\vi$ where $\rho_i$ is the ion
mass density and $d_t \equiv d/dt \equiv \partial_t + \vi \cdot \grad$ is
the convective derivative), is retained in the momentum equation for
the ionised component but its effect on Ohm's
law is ignored in BB94.  This limits the applicability of the results
to the ambipolar diffusion limit.

To better appreciate this point, let us briefly recast the
two fluid formulation of BB94 starting with seperate ion and electron
fluid equations.  The equation of motion for the ionised fluid  is 
derived assuming that magnetic field is frozen in the
electron fluid,
\begin{equation}
     0 = -e\,n_e \left(\E + \ve\cross\B/c\right)
     \label{eq:elmotion}
\end{equation}
and summing the electron and ion momentum equations, to
yield
\begin{equation}
     \frac{d \vi}{dt} + \frac{\grad P_i}{\rho_i} + \grad \Phi + \nu_{in}\,\left(\vi - \vn \right)=
     \frac{\J\cross\B}{c\,\rho_i}
     \label{eq:ion_mom}
\end{equation}
equation (15) of BB94.  Here
$e$ is the electronic charge, $n_e$ is the electron number density, $\nu_{in}$ is the ion-neutral collision frequency,  
$\ve, \,\vi$ and $\vn$ are the electron, ion and neutral velocities, $\Phi$ is the gravitational potential, $P_i$ is the ion 
pressure, $\E,\,\B$ are the electric and magnetic fields and
$c$ is the speed of light and $\J =e\,n_e (\vi - \ve)$ is the current density.
Taking the curl of
(\ref{eq:elmotion}) and using Maxwell's equation, $ c \,\curl
\E = - \partial_t \B$, will give $\partial_t\,\B =
\grad\cross\left(\ve\cross\B\right)$, i.e. the magnetic field is convected
away by the electron fluid.  If we want to express the right hand side of
induction equation in terms of ion velocity, we obtain
\begin{equation}
     \delt \B = \grad\cross\left(\vi\cross\B\right) - \curl\FH \,,
     \label{eq:indn}
\end{equation}
where the Hall term is
\begin{equation}
     \FH =\frac{\J\cross\B}{e\,n_e}
     \label{eq:FH}
\end{equation}

Since $\J\cross\B/c \sim \rho_i\,\left(d_t\,\vi +
\nu_{in}\,\vi\right)$ (here $\nu_{in}$ is the ion-neutral collision
frequency), the Hall term can be dropped
from the induction equation only if $\left(\nu_{in}, \omega \right)
\ll \omega_{ci} \left(= e\,B/(m_i\,c)\right)$, i.e. the ion-gyration
period ($\omega_{ci}^{-1}$) is smaller/ faster than the dynamical time ($\omega^{-1}$) or
the ion-neutral collision time ($\nu_{in}^{-1}$).
We see that unlike a two component
electron-ion plasma, where the Hall term can be introduced only through
the ion inertial term ($d_t\,\vi$), in a weakly ionized
multi-component plasma, the Hall effect appears either via
ion-neutral collision or via the $d_t\,\vi$ term or both.

Replacing $\partial_t\,\B$ by $\Delta B/\delta t$ and $\curl\FH$ by
$c\,B\,\Delta B/(4\, \pi\,n_i\e\L\,\Delta x),$ one sees that the Hall term
scales as $1/(\omega_{pi}\,L)$, (here $\omega_{pi} =
\left(4\,\pi\,e^2\,n_i/m_i\right)^{0.5}$ is the ion plasma frequency and 
$L$ is the characteristic size of the system), i.e. the Hall term is important 
on a scale shorter than the ion-inertial scale.
Clearly, Hall MHD introduces two disparate, interacting scales, a microscopic scale, i.e. 
the ion-skin depth ($\delta_i = c/\omega_{pi} \equiv V_A/\omega_{ci}$) and a macroscopic scale,
the disc size.  The Hall term can be dropped if ion-inertial effects
are unimportant. Leaving out the effect of inertia in the induction
equation but retaining them in dynamics is not consistent and, in such a scenario, one would expect 
that \mri will merely shift towards long wavelength, as has already been noted by BB94.

If we start with the ion
equation of motion (BB94 (15)), and express the electric field as
\bq
\E = - \frac{\vi\cross \B}{c} + \frac{m_i}{e} \left[ \frac{d\vi}{dt} +
\grad \Phi +\frac{\grad P_i}{\rho_i} + \nu_{in}\,\,\bmath{v}_D \right]
\label{eq:efi}
\eq where $\bmath{v}_D = \vi - \vn$, then
taking the curl and using Maxwell's equation, one arrives at the
following induction equation
\bq
\delt \B = \curl\left(\vi\cross\B\right) - \frac{m_i\,c}{e}\left[
\nu_{in}\,\curl\bmath{v}_D + \curl\frac{d\vi}{dt} \label{eq:indn} 
\right].  \eq 
Here uniform density is assumed while operating with the curl on equation (\ref{eq:efi}).
This
equation has an additional term in comparision with equation (16) of
BB94, with important consequences on the magnetic diffusivity, since the rate of
change of magnetic flux is given as
\begin{eqnarray}
\frac{d}{dt}\int \int \B\cdot\ds = \int\int \delt{\B} \cdot \ds + 
\oint \vi\cross \bmath{dl} \cdot \B \nonumber\\
\equiv
\int \int \left[ \delt{\B} - \curl\left(\vi\cross\B\right)\right]\cdot \ds.
\end{eqnarray}
Making use of equation (6) in (7), we get
\begin{eqnarray}
\frac{d}{dt}\int\int\hat{\B}\cdot\ds = - \frac{1}{\beta_i} \oint 
\bmath{v}_D \cdot \bmath{dl} \nonumber \\
- \frac{1}{\omega_{ci}} \int \int \curl\left[\delt{\vi} - 
\vi\cross\left(\curl\vi\right)\right]\cdot \ds
\end{eqnarray}
Here $\hat{\B} = \B/B$ and use has been made of $d\vi/dt = \partial_t 
\vi - \vi\cross\left(\curl\vi\right)+\grad\vi^2/2$.
The ion Hall parameter $\beta_i = \omega_{ci}/{\nu_{in}}$ gives the ratio 
between the ion-cyclotron to ion-neutral collision frequencies.
The above equation can be rewritten as
\bq
\frac{d}{dt}\int\int \left[ \hat{\B} + 
\frac{1}{\omega_{ci}}\,\curl\vi \right]\cdot \ds
= - \frac{1}{\beta_i} \oint \bmath{v}_D \cdot \bmath{dl}.
\label{eq:flux_transport}
\eq We see from equation (\ref{eq:flux_transport}) that the
generalized flux that is a combination of magnetic flux and vorticity
is not conserved.  The rate at which this flux decays is directly
related to the collisional coupling between ions and neutrals.  If the
ion magnetization is weak, i.e. the ion-cyclotron frequency is less than
ion-neutral collision frequency ($\beta_i \rightarrow 0$), then the
flux-decay rate could be very large for a finite ion-neutral drift
speed $\bmath{v}_D$.  However, if the relative ion-neutral drift is
negligible, the generalized flux is conserved irrespective of the ion
magnetization level.  Thus it is the combination of the magnetic flux
and the vorticity that is conserved in the absence of collision
($\beta_i \rightarrow \infty$). The BB94 formulation assumes that
magnetic flux is frozen in the ion-fluid which is valid if apart from
ignoring the right hand side of equation (\ref{eq:flux_transport}), we also assume that
$\omega_{ci} \rightarrow \infty$.  In this limit however, the role of
ion inertia becomes increasingly unimportant and we approach W99
limit.  Clearly, BB94 does not treat the effect of ion inertia in a
consistent fashion and their results are not applicable in most of the
weakly ionized parameter space where the ion Hall parameter is $\sim 1$.
In Fig.1, we plot the range of applicability of BB94 and W99.  We see
that BB94 is applicable when $\beta_i \gg 1$ for $\omega_{ci}>\Omega$
for arbitrary relation between $\nu_{in}$ and $\Omega$.  BT01 show that $\beta_i
\gg 1$ implies that the Hall term dominates over ambipolar term.
W99 is applicable for $\omega_{ci}> \Omega$ and $\nu_{in} > \Omega$
for arbitrary $\beta_i$. Therefore, $\omega_{ci}/\Omega < 1$ and
$\nu_{in}/\Omega < 1$ is an unexplored region in BB94 and W99
framework.  We see that in a PPD, for a milliGauss field, for a
positive grains of mass $10^{-12}-10^{-15}\, \mbox{g}$,
$\omega_{ci}/\Omega < 1$. As has been noted elsewhere, near the mid-plane of PPDs, dust grains can be the 
dominant charged constituent over extended regions ($\sim 1 - 5$ AU). For sub-micron sized grains
$0.1\,\mu\mbox{m}$, negatively charged grains dominate whenever $n_n \gtrsim 10^{11}\,\mbox{cm}^{-3}$ and
positively charged grains dominate for $n_n \gtrsim 10^{14}\,\mbox{cm}^{-3}$ \citep{w5}. Depending upon 
neutral density in the disc,
the ratio $\nu_{in} / \Omega$ can have any value and thus, it is
importnat to extend the BB94 analysis to the unexplored regions
with full, Hall and ambipolar effects in the spirit of W99.
\begin{figure}
      \includegraphics[scale=0.45]{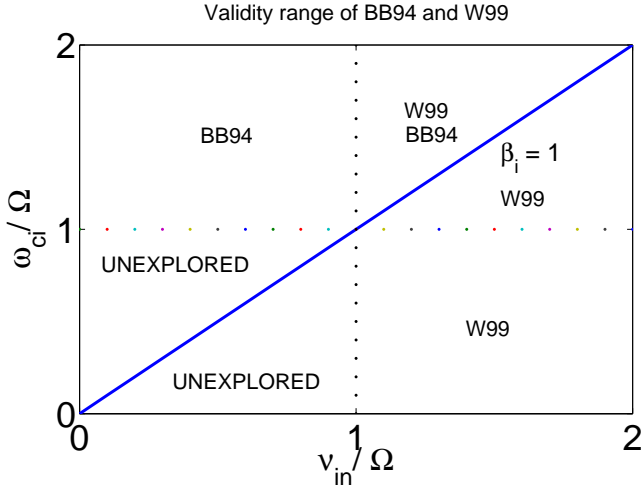}
      \caption{\large{The region of applicability of BB94 and W99.}}
\end{figure}
We shall give a general derivation of the induction equation in a 
weakly ionized medium in the text though, we adopt the conductivity
approach of W99 to investigate the effect of ion inertia on the \mri 
and find a significant increase in the growth rate. Further, the parameter window
in which instability operates expands considerably to large scale 
lengths. It will have an important bearing on the angular
momentum transport and onset of turbulence.
\section{Formulation - MHD equations}
The dynamics of a weakly ionized disc, consisting of electrons, ions, 
neutrals and charged and neutral dust grains,
in the presence of a gravitational field $\Phi$ of a central mass 
point $M$ is described by a set of multifluid equations.
A multi-fluid approach representing each and every particle species 
is not very fruitful since
depending upon the fractional ionization, the presence of some of the 
ionized component in the disc can be neglected. We shall assume
that the ion density in the disc is mainly due to the presence of 
positively charged grains. Such a situation will correspond to a 
very dense region of PPDs \citep{w5}.

As the disc matter is weakly ionized, generally the inertia of the 
charged species are
neglected (W99). However, apart from pure scientific curiosity about 
the effect of charged particle inertia on the
\mri, there are astrophysical environments where the plasma inertial 
effects may become important. For example,
ion inertial effect may compete with collisional and electromagnetic 
effects in the dense region of PPDs
as well as in the high fractional ionization discs around magnetic 
cataclysmic variables. An estimate of the ion inertial, collisional and electromagnetic effects are given below. 
Since we wish to investigate the role of ion dynamics
on the \mri, we shall retain the inertia term in the ion equation of motion.
This will result in conductivity tensor ${\bf{\sigma}}$ becoming frequency dependent.

The basic set of equations describing a partially ionized, 
non-self-gravitating magnetized plasma disc,
consisting of electrons, ions, and neutrals, are as follows.
The continuity equation is
\bq
\frac{\partial \rho_j}{\partial t} + \grad\cdot\left(\rho_j\,\vj\right) = 0.
\label{cont}
\eq
Here $\rho_j$ is the mass density and $\vj$ is the velocity of the 
various plasma components.

The momentum equations for electrons, ions and neutrals are

\bq
0=  - e\,n_e\,\left(\E' + \frac{\ve\cross \B}{c}\right) - 
\rho_e\,\nu_{en}\,\ve,
\label{em1}
\eq
\bq
\rho_i\,d_t\,\vi= e\,n_i\,\left(\E' + \frac{\vi\cross \B}{c}\right) - 
\rho_i\,\nu_{in}\,\vi - \rho_i\,d_t\,\vn,
\label{im1}
\eq
\bq
\rho_n\,d_t\,\vn = - \grad{P} - \rho_n\,\grad\Phi + 
\displaystyle\sum_{e, i} \rho_j\,\nu_{j n}\,\vj.
\label{nm1}
\eq
Here $\vj$ is the drift velocity through the neutrals, $\Fj = 
q_j\,n_j\,\left(\E' + \vj\cross \B/c\right)$ is the
Lorentz force, $n_j$ is the number density, and,  $j$ stands for 
electrons ($q_e = -e$),
and ions ($q_i = e$). Grains are assumed to have single positive 
electic charge. The electric field $\Ep = \E + \vn \cross \B$
is written in the frame comoving with the neutrals. In equation 
(\ref{nm1}), the gravitational potential $\Phi$ due to central
mass, is given by
\bq
\Phi = - \frac{G\,M}{\sqrt{r^2 + z^2}}.
\eq
At the disc midplane, the Keplerian centripetal acceleration $v_K^2/r 
\equiv G\,M/r^2$ balances the radial
component of the gravitational potential. Equations 
(\ref{em1})-(\ref{im1}) does not have a pressure gradient term since pressure
effects will be negligible in a  weakly ionized disc. The effects
of ionization and recombination are also omitted from the neutral 
dynamics for the same reason.

The collision frequency is
\bq
\nu_{j n} = \gamma_{j n}\,\rho_n = \frac{\sigv_j}{m_n + m_j}\,\rho_n.
\label{cf0}
\eq
Here $\sigv_j$ is the rate coefficient for the momentum transfer by 
the collision of the $j^{\mbox{th}}$ particle
with the neutrals. The ion-neutral
and electron-neutral rate coefficients are \citep{d6}

\begin{eqnarray}
<\sigma\,v>_{in} = 1.9 \cross 10^{-9}\quad 
\mbox{cm}^3\,\mbox{s}^{-1}\nonumber \\
<\sigma\,v>_{en} = 4.5 \cross 
10^{-9}\,\left(\frac{T}{30\,\mbox{K}}\right)^{\frac{1}{2}}\quad 
\mbox{cm}^3\,\mbox{s}^{-1}.
\label{cf1}
\end{eqnarray}
Adopting a value of $m_i = 30\,m_p$ for ion mass and $m_n = 2.33\,m_p$ for mean neutral mass where 
$m_p = 1.67\times 10^{-24}\,\mbox{g}$ is the proton mass, the
ion neutral collision frequency can be written as
\bq
\nu_{in} = \rho_n\,\gamma \equiv \frac{m_n\,n_n\,<\sigma\,v>_{in}}{m_i + m_n}
= 1.4\cross 10^{-10}\,n_n\,\mbox{s}^{-1}.
\label{cf2}
\eq This also gives the limiting value for very small grains ($\sim 3 - 3000\,\AA$).  For
larger, micron sized grains, ion-neutral collision rate can vary
between $10^{-10}$ to $10^{-5}$ for sizes ranging between a few Angstrom to a few microns. 
This can be seen if we write the collision rate as
\citep{nu}
\begin{equation}
<\sigma\,v>_{in} = 2.8 \cross 10^{-5}\, T_{30}^{\frac{1}{2}}\,a_{-5}^{2},
\label{cg}
\end{equation}
where $T_{30}$ is the gas temperature and $a_{-5}$ is the grain radius
in units of 30\,K and $10^{-5}$\,cm respectively.

We shall rewrite equations (\ref{cont})-(\ref{nm1}) in the local 
Keplerian frame. Thus, velocity $\v$ represents the departure from
the Keplerian motion; the fluid velocity in the laboratory frame is 
$\v + \vk$ and $\partial_t$
is $\partial_t + \Omega \,\partial_{\phi}$ in the laboratory frame, 
where $\vk = \sqrt{G\,M/r}\,
\fh$ is the Keplerian velocity in the canonical cylindrical 
coordinate system $(r, \phi, z)$.

Noting that near the disc midplane, on scales small compared with the 
disc thickness, the radial gradient
in gravitational potential will be exactly cancelled by the 
centripetal term due to Keplerian motion, and,
$(r, \phi)$ component of the equation (\ref{im1})-(\ref{nm1}) in the 
absence of any tidal effects, can be rewritten as,
\bq
\A\vi= \frac{e}{m_i}\,\left(\E' + \frac{\vi\cross \B}{c}\right) - 
\nu_{in}\,\vi - \A\vn,
\label{im2}
\eq
\bq
\A\vn = - \frac{\grad{P}_n}{\rho_n} + \frac{\J\cross\B}{c\,\rho_n} + 
{\large O}\left(\frac{\rho_i}{\rho_n}\right),
\label{nm2}
\eq
where operator
$\A = \left(
\begin{array}{cc} d_t   & -2\,\Omega \\
                  0.5\,\Omega  & d_t
  \end{array}
\right)$. The induction equation can be written as
\bq
\partial_t\,\B = \curl \left(\v\cross\B \right) - c\,\curl\Ep - 
1.5\,\Omega\,B_r\,\fh.
\label{ind1}
\eq
In (\ref{ind1}), $\curl\Ep$ contains the effect of non-ideal MHD
and the last term accounts for the generation of the toroidal field from
the poloidal one due to differential rotation of the disc (W99).

Before describing the conductivity approach and giving a formulation
to the problem at hand, we shall give a general derivation of the 
induction equation starting from equation (\ref{em1}) and estimate the
range of applicability of the ion-inertial effects in the accretion
discs.  Writing $\E' = - \ve\cross \B/c + \eta\,\J$ as
\bq
\E' = - \frac{\vi\cross \B}{c} + \eta\,\J + \FH.
\eq
Here
\bq
\eta = \frac{c^2}{4 \pi}\frac{m_e\,\nu_{en}}{n_e\,e^2} \equiv 
\frac{c^2}{4 \pi}\frac{m_e\,n_n\,<\sigma\,\,v>_{en}}{n_e\,e^2}
\eq
is the electrical resistivity of the gas and $\FH = \J\cross\B/e\,n_e$
is the Hall term.  Even in the absence of ion inertia - the so called
strong coupling approximation, \citep{shu}, the ambipolar term
modifies the induction equation, i.e. when $\rho_i\,\nu_{in}\,\vi =
\J\cross \B/c$,
\bq
\E' = \eta\,\J + \FH - 
\frac{\left(\J\cross\B\right)\cross\B}{c\,\rho_i\,\nu_{in}}.
\eq
When ion inertial terms are also present,
\bq
\vi = \frac{\J\cross\B}{c\,\rho_i\,\nu_{in}} - 
\frac{1}{\nu_{in}}\frac{d\vi}{dt} + {\large 
O}\left(\frac{\rho_e}{\rho_n}, \frac{\rho_i}{\rho_n}\right)
\eq
and, the generalized Ohm's law becomes
\bq
\E' = \eta\,\J + \FH - \left[ 
\frac{\left(\J\cross\B/c\right)}{\rho_i\,\nu_{in}}
+ \A_1\vi \right]\cross\B,
\label{ef}
\eq
where
$\A_1 = \nu_{in}^{-1}\A $.
Now taking the curl of equation (\ref{ef}), one may write the generalized 
induction equation as
\begin{eqnarray}
\frac{\partial \B}{\partial t} = \curl\left[ \v\cross\B - 
\frac{4\,\pi\,\eta\,\J}{c}- \frac{\J\cross\B}{e\,n_e}
\right]
\nonumber \\
+ \curl\left[\frac{ 
\left(\J\cross\B\right)\cross\B}{c\,\rho_i\,\nu_{in}} - 
(\A_1\vi)\cross\B \right]
.
\label{ind2}
\end{eqnarray}
The induction equation (\ref{ind2}) on the right hand side contains
inductive, Ohmic diffusion, Hall, ambipolar, and
ion inertial terms respectively.  The set of equation (\ref{cont}),
(\ref{im1}), (\ref{nm1}) and (\ref{ind2}) can be closed by an equation
of state.

Assuming that ion-inertial time scale is of the order of disc
rotation period, we may write $\A_1 \sim \Omega/\nu_{in}$.  Then 
the ratio of the ion inertial term to the inductive term can be written as 
\bq
\frac{|\v\cross\B|}{|\A_1\vi\cross\B|} \sim \frac{\nu_{in}}{\Omega}.
\label{RE1}
\eq
A different ratio that measures the coupling of the
neutral to the ion with respect to the Keplerian frequency have been
used by BB94 \citep{MN}
\bq
Re_A \equiv \frac{\nu_{ni}}{\Omega} = \alpha\,\frac{\nu_{in}}{\Omega},
\label{REA}
\eq where $\alpha = \rho_i/\rho_n$.  It is clear from equation
(\ref{REA}) that \mri can act on both the neutral as well as on the ion
fluid simultaneously if $\alpha \sim 1$.



Assuming an equation of state or dropping the pressure gradient term 
in the neutral equation of motion, equations (\ref{cont}), (\ref{im2}), (\ref{nm2}) and 
(\ref{ind2}) with Maxwell's equations
\bq
\grad\cross \B = \frac{4\,\pi}{c}\J,\quad \grad \cdot \B = 0,
\label{aml}
\eq
form a complete set.
\section{Linearization}
We consider a thin disc implying that the radial scale over which 
physical quantities vary is much
larger than the disc scale height, $H = C_s/\Omega$. The initial steady state is assumed uniform and
homogeneous with a vertical magnetic field $\B = B \zh$ and zero $\v, 
\vn, \grad P, \E'$ and $\J$.

We shall assume transverse fluctuations  and denote resulting 
two-dimensional vectors by subscript $\perp$,
to investigate \alf modes in the disc.  We seek plane wave solution of the 
form $\exp i \left(\omega\,t - k\,z\right)$.

\subsection{The conductivity tensor}
\label{sec:Basic Model}
The conductivity tensor \boldmath$\sigma$\unboldmath\ can be found by 
considering the drifts
of charged particles in response to the electromagnetic field 
\citep{c,nh,nu,w5}. We shall first derive
conductivity tensor from equation of motion for charged particles,
from (\ref{em1}) and (\ref{im2}) by eliminating $\ve$ and $\vi$ in 
favour of $\E'$ and $\B$ and then 
give the expressions for parallel, Hall and Pedersen component of this tensor 
${\bf{\sigma}}$ in the generalized Ohm's law, $\J = {\bmath\sigma} \cdot \Ep$.

Since we assume a homogeneous steady state, it implies that the inohomegenity scale length 
(in this case verticle scale height H of the disk) is much larger than the characteristice wavelength of 
the normal modes in the disc. Thus, the contribution of the pressure gradient will be neglected while 
inverting equation (\ref{nm1}). Furthermore, we shall also ignore terms  of the order $\sim \rho_i/\rho_n$ in the conductivity 
tensor. Expressing velocities $\vj$ in
equations (\ref{em1}) and (\ref{im2}) in terms of electric field $\E'$ 
the relationship between $\J$ and $\E'$ can be written as,
\bq
	\J = \mbox{\boldmath$\sigma$}\cdot \E' = \sigma_{\parallel} \Epa +
	\sigma_\mathrm{H} \Bh \cross \Epe + \sigma_\mathrm{P} \Epe  \,,
\label{olaw}
\eq
where $\sigma_\parallel$, $\sigma_\mathrm{H}$ and $\sigma_\mathrm{P}$
are the field-parallel, Hall and Pedersen components of the
conductivity tensor \boldmath$\sigma$\unboldmath\ and $\Bh$ is the 
unit vector along
the magnetic field. If charged species $j$
has particle mass $m_j$, charge $Z_je$, number density $n_j$, then the Hall
parameter is given as
\bq
\beta_j= \frac{Z_jeB }{ m_j c} \, \frac{m+m_j}{ <\sigma v>_j \rho}\,
\label{hallp}
\eq
where $m $ is the mean neutral particle mass and have dropped 
subscript \`{}n\'{} from neutral quantities. As noted in section 2, the Hall
parameter determines the magnitude of the magnetic flux transport. 
The ratio of ion to electron
Hall parameter suggest that in the protostellar discs, $\beta_i/\beta_e \sim 10^{-3} \ll 1$.
Recall that BB94 formulation is valid when the ion-Hall parameter is large ($\beta_i \gg 1$), i.e. the when Lorentz 
force dominates the ion-neutral collisional momentum exchange. However, this limit implies strongly
magnetized ions and infrequent collisions.

The conductivity tensor is frequency dependent in the presence of ion inertia. The parallel conductivity
is
\bq
\sigma_{\parallel} = \frac{e\,c\,n_i}{B} \left[ \beta_e + 
\frac{\beta_i}{1+ \left(\frac{\omega}{\nu_{in}}\right)^2}
- i \, \beta_i\,\frac{\frac{\omega}{\nu_{in}}}{1+ 
\left(\frac{\omega}{\nu_{in}}\right)^2}\right].
\label{sigma0}
\eq
Since the plasma is quasi-neutral, we have assumed $n_e = n_i$. The 
conductivity has become complex. In the low frequency
limit, when ion inertial effect is unimportant, i.e. $\omega/\nu_{in} 
\rightarrow 0$, $\sigma_{\parallel}$ reduces
to W99. In $\omega/\nu_{in} \rightarrow \infty$, $\sigma_{\parallel}$ 
has both a real and an imaginary
component
\bq
Re[\sigma_{\parallel}]  \simeq \frac{e\,c\,n_i}{B} \, \beta_e,
\label{sigma0R}
\eq
and,
\bq
Im[\sigma_{\parallel}] \simeq - \frac{e\,c\,n_i}{B}\, 
\left(\frac{\omega_{ci}}{\omega}\right).
\label{sigma0I}
\eq

The Pedersen conductivity is
\bq
\sigma_{P} = \sigma_P^0 \left[ 1 + \frac{\Delta \sigma_P}{\sigma_P^0}\right].
\label{PCT}
\eq
Here frequency independent part, $\sigma_{P}^{0}$ is
\bq
\sigma_{P}^{0} = \frac{e\,c}{B}\displaystyle\sum_j 
\frac{n_j\,Z_j\,\beta_j}{1+\beta_j^2}\,,
\label{PCD}
\eq
and frequency dependent part, $\Delta \sigma_P$ is
\bq
\Delta \sigma_P = 
\frac{e\,c}{B}\left(\frac{n_i\,\beta_i}{1+\beta_i^2}\right) 
(Q(\omega) - 1).
\label{PCAC}
\eq
Here $Q(\omega) =  (1 + \beta_i^2)\, D_1/D_2$, $D_1= i\,\omega/\nu_{in} + 1$, 
$D_2= D_1^2 + \hat\Omega_1\,\hat\Omega_2$, $\hat\Omega_1=2\,\Omh + \beta_i$, 
$\hat\Omega_2= 0.5\,\Omh + \beta_i$ and $\hat{\Omega} = \Omega/\nu_{in}$. In 
order to isolate real and imaginary part of $\sigma_P$ and
investigate the low and high frequency limits, we write $Re[Q(\omega)]$ 
and $Im[Q(\omega)]$as
\bq
\frac{Re[Q(\omega)]}{1+\beta_i^2} = \frac{1 + \left(\frac{\omega}{\nu_{in}}\right)^2 + 
\hat\Omega_1\,\hat\Omega_2}
{\left[1 - \left(\frac{\omega}{\nu_{in}}\right)^2 + 
\hat\Omega_1\,\hat\Omega_2\right]^2 + 
4\,\left(\frac{\omega}{\nu_{in}}\right)^2
},
\label{req}
\eq
and,
\bq
\frac{Im[Q(\omega)]}{1+\beta_i^2} = - \frac{\left(\frac{\omega}{\nu_{in}}\right)\left[1 + 
\left(\frac{\omega}{\nu_{in}}\right)^2 - \hat\Omega_1\,\hat\Omega_2\right]}
{\left[1 - \left(\frac{\omega}{\nu_{in}}\right)^2 + 
\hat\Omega_1\,\hat\Omega_2\right]^2 + 
4\,\left(\frac{\omega}{\nu_{in}}\right)^2
}
\label{imq}
\eq
In the low frequency limit, $\omega/\nu_{in} \rightarrow 0$, and assuming 
$\omega \sim \Omega$, $Re[Q(\omega)] \simeq 1$ and $Im[Q(\omega)] \simeq 0$.
Thus $\sigma_{P} = \sigma_{P}^{0}$. In the high frequency limit, when 
$\omega/\nu_{in} \rightarrow \infty$, $Re[Q(\omega)] \approx 0$ and
$Im[Q(\omega)] \approx - \nu_{in}/\omega$. Thus,
\bq
Re[\sigma_{P}] = \frac{e\,c}{B} \frac{n_e\,\beta_e}{1+\beta_e^2},
\eq
and,
\bq
Im[\sigma_{P}] = - \frac{e\,c}{B} 
n_e\,\beta_i\,\left(\frac{\omega}{\nu_{in}}\right)^{-1}.
\eq
Like the parallel conductivity, the real part of the Pedersen conductivity in the high frequency 
limit is mainly
due to electron magnetization and imaginary part is due to ion magnetization. Complex resistivity is well known
in LCR circuits where the resonance condition is found by setting 
imaginary part of the impedance to zero. In the present case, a similar resonance 
condition can be derived by setting numerator of equation (\ref{imq}) to zero, i.e. $\omega^2 / \nu_{in}^2 = 
-1 + \hat\Omega_1\,\hat\Omega_2$,
we get
\bq
  1 + \frac{\Delta \sigma_P}{\sigma_P^0} \simeq  \frac{\hat\Omega_1\,\hat\Omega_2}{2\,\left( 1 + 
\frac{\omega^2}{\nu_{in}^2}\right)}.
\label{scg}
\eq
It is important to note that the scale of frequency dependent conductivity associated with the resonance 
($i\,\omega \sim \Omega$), can become larger than the dc conductivity. We anticipate, therefore, that the 
frequency dependent Pedersen conductivity will significantly modify the \mri.

The Hall conductivity is
\bq
\sigma_H = - \left( \begin{array}{cc}
        1 & \frac{\Delta \sigma_{Hr}}{\sigma_{H}^{0}} \\
        \frac{\Delta \sigma_{H \phi}}{\sigma_{H}^{0}} & 1
       \end{array}
\right)\left( \begin{array}{c}
        \sigma_{H}^{0} \\
        \sigma_{H}^{0}
       \end{array}
\right)
\eq
where the dc part is given as
\bq
\sigma_H^0 =  \frac{e\,c\,n_i}{B}\sum_j \frac{Z_j}{1+\beta_j^2}\,.
\eq
The frequency dependent parts, $\Delta \sigma_{Hr}$, and, 
$\sigma_{H\phi}$ are
\begin{eqnarray}
\Delta \sigma_{Hr} = \frac{e\,c}{B}\,\frac{n_i\beta_i^2}{1+\beta_i^2} 
(H_1(\omega) - 1)\,, \nonumber\\
\Delta \sigma_{H\phi} = 
\frac{e\,c}{B}\,\frac{n_i\beta_i^2}{1+\beta_i^2} (H_2(\omega) - 1)\,,
\end{eqnarray}
and,
\bq
H_j(\omega) = \frac{\left(1+\beta_i^2\right)\,\hat\Omega_j}{\beta_i\,D_2},
\label{defH}
\eq
for $j = 1, 2$. The radial and azimuthal component of the Hall conductivities 
are not equal due
to the unequal radial and azimuthal coefficient in the ion momentum 
equation. The real and
imaginary part of $H_j$ is given by equations (\ref{req})- 
(\ref{imq}) if we recognize that right hand side
of equation (\ref{defH}) has a factor $1/D_2 = Q(\omega)/D_1\,(1 + \beta_i^2)$. Thus, 
near resonance
\bq
\left( 1 + \frac{\Delta \sigma_{Hr}}{\sigma_{H}^{0}}\right) \simeq \frac{\hat\Omega_1\,\left( 1 +
\hat\Omega_1\,\hat\Omega_2\right)}{\beta_i\left[\left( 1 +
\hat\Omega_1\,\hat\Omega_2\right)^2 + 4\,\left(\frac{\omega}{\nu_{in}}\right)^2\right]}.
\label{chr}
\eq
Except for $\hat\Omega_1$ becoming $\hat\Omega_2$, the remainder of the expression 
for the azimuthal factor  will be identical to equation (\ref{chr}). 

The analogy to LCR resonance can be brought closer if we express Ohm's law, equation (\ref{olaw}) in 
diagonal form. To that end, we shall express $\Epe$ in the eigen-basis vectors of the rotation operator
$\bmath{\hat e}_{\pm} = (\bmath{e}_x \pm i\,\bmath{e}_y)/\sqrt{2}$. Then $\Bh\cross \bmath{\hat e}_{\pm} = \mp i\,\bmath{\hat e}_{\mp}$ 
and, Ohm's law for the transverse component $\J_{\pm}$ can be written as
\bq
\J_{\pm} = \left(\sigma_P \mp i\,\sigma_H \right)\,\Epe_{\pm}.
\label{om1}
\eq
While writing equation (\ref{om1}), we have assumed $\sigma_{Hr} = \sigma_{H\phi} = \sigma_{H}$. We may write 
$\J_{\pm} = Z_{\pm}\left(\omega\right)\,\Epe_{\pm}$. Here the impedance 
$Z_{\pm}\left(\omega\right) = R_{\pm} + i\,X_{\pm}\left(\omega\right)$, with 
$R_{\pm} = Re[\sigma_{p}] \pm Im[\sigma_{H}]$, and, $X_{\pm} = Im[\sigma_{p}] \mp Re[\sigma_{H}]$. Near 
resonance, $Z_{\pm} = R_{\pm}$. Thus 
\bq
\J_{\pm} =  R_{\pm} \Epe_{\pm} \propto \frac{\Epe_{\pm}}{\left(1 + \frac{i\,\omega}{\nu_{in}}\right)\left(1 - \frac{i\,\omega}{\nu_{in}}\right)}.
\eq
For a growing mode $i\,\omega \sim \Omega$, and, oscillations in the current can take place in the absence of a 
neutral-frame electric field $\Epe \rightarrow 0$. In a weakly ionized disc, if $\beta_i \lesssim 1$, the 
oscillation in the current is set by the ions diffusing across the ambient magnetic field. For $\nu_{in} \sim i\,\omega$, near resonance, collision will act like a driver of the resonance.

We note that near resonance, the conductivity may change sign. From (\ref{scg}), when $\Omega < \nu_{in}$, the 
negative conductivity will play an important role. The ratio of the dynamical to the ion-neutral collision frequency 
determines whether negative conductivity is important.  The DC conductivity becoming negative within 
certain frequency range, in the microwave irradiation is well known in the condensed matter literature, 
e.g. \cite{RY}.   
\subsection{Dispersion relation}
The linearized neutral equation of motion (\ref{nm2}) can be written as
\bq
\left(
\begin{array}{cc}\omega    & 2\,i\,\Omega\\
                   \frac{- i\,\Omega}{2} &\omega
  \end{array}
\right)\dvp
  = - k\,v_A^2\,\left(\frac{\dBp}{B}\right),
\eq
where subscript $\perp$ denotes two dimensional vector in the disc plane and $v_A = B/\sqrt{4\,\pi\,\rho}$ is 
the \alf velocity in the total fluid.
The linearized induction equation, after substituting for $\dvp$ is given as
\begin{eqnarray}
\left( \begin{array}{cc}
\omega_{A}^2+3\Omega^2  & 2 i\omega \Omega \\
-2 i\omega \Omega     &\omega_{A}^2
\end{array}
\right) \dBp =
ik\,c \left( \begin{array}{cc}2\Omega&i\,\omega\\-i\omega&\frac{\Omega}{2}
\end{array}\right)
  \dEp,
\label{lind}
\end{eqnarray}
where $\omega_{A}^2 = \omega^2 - k^2\,v_{A}^2$.

In the ideal MHD limit, when $\dEp = 0$, one recovers 
magnetorotational mode. In the
absence of rotation, a dispersion relation for ideal MHD can be 
derived by setting determinant
of left hand side matrix to zero. The departure from ideal MHD is due 
to the collisional effects.
They will appear when electric field is eliminated in favour of 
magnetic field utilizing Ohm's
law and Maxwell's equation.

Making use of linearized Ampere's law
\bq
\dJp = \frac{i\,k\,c}{4\,\pi}
      \left( \begin{array}{cc}
                                  0&1\\
                                  -1&0
              \end{array}
       \right)\dBp,
\eq
in the generalized Ohm's law,
\bq
\dEp = \frac{-i\,k\,c}{4\,\pi\,\Delta(\omega)}
      \left( \begin{array}{cc}
                                  s\,\sigma_{Hr}&-\sigma_P\\
                                  \sigma_P&s\,\sigma_{H\phi}
              \end{array}
       \right)\dBp,
\eq
where $\Delta(\omega)= \sigma_{Hr}(\omega)\,\sigma_{H\phi}(\omega)+\sigma_P(\omega)^2$ 
and, $s = sign(B_z)$. Introducing normalized variable
$y = i\,\omega/\Omega$ frequency dependent part of the
conductivities can be written as,
\begin{eqnarray}
Q(y) = (1+\beta_i^2)\,F(y)\,,\nonumber\\
H_1(y) = \frac{(1+\beta_i^2)\,\hat\Omega_1\,F(y)}{D_1\,\beta_i}\,,\nonumber\\
H_2(y) = \frac{(1+\beta_i^2)\,\hat\Omega_2\,F(y)}{D_1\,\beta_i}\,,
\end{eqnarray}
where $F(y) = D_1(y)/D_2(y)$ and $D_1(y) = 1 + \hat{\Omega}\,y\,, D_2 
= D_1^2 + \hat{\Omega}_1\,\hat{\Omega}_2,\,
y = i\,\omega/\Omega$.

Eliminating
$\dEp$ from equation (\ref{lind}), one gets the
following dispersion relation
\bq
a\,\left(\frac{k\,v_A}{\Omega}\right)^4 + b 
\left(\frac{k\,v_A}{\Omega}\right)^2 + c =0
\label{drf}
\eq
\begin{eqnarray}
a=\chi^2\,G(y)^2+ \chi\, G(y)\, \left(2\,s\sigma_{H\phi}+0.5\,s\,\sigma_{Hr}+2\,y\,\sigma_P\right) \sigmaperp^{-1}
\nonumber \\
+ \left(y^2 +1 \right)\,G(y),
\label{qa}
\end{eqnarray}
\begin{eqnarray}
b= \left(2\,y\,\sigma_P-1.5\,s\,\sigma_{H\phi}\right)(y^2+1)\,\chi\,G(y)\,\sigmaperp^{-1}\nonumber\\
+\chi^2\,G(y)^2\,(2\,y^2-3),
\label{qb}
\end{eqnarray}
\bq
c=\chi^2\,G(y)^2\,y^2\,(y^2+1).
\label{qc}
\eq
Here $G(y) = \Delta(y)/\sigmaperp^2$, $\sigmaperp =\sqrt{\sigma_p^0{}^{2} + \sigma_H^0{}^{2}}$ and the 
parameter $\chi = \omega_c/\Omega$ is the
normalized critical frequency $\omega_c = 4\,\pi 
\left(v_A/c\right)^2\,\sigmaperp$ similar to W99. As has been noted in W99, the 
ideal MHD description is valid in the
large $\chi$ limit. When $k\,v_A \geq \omega_c$ non-ideal MHD effect 
become important.

When ion inertial effects are ignored, the expressions for $a, \,b \,\mbox{and}\, c$, in equations 
(\ref{qa})-(\ref{qc}) reduces to W99. Ideal MHD is recovered in
$\chi \rightarrow \infty$ limit. As has been discussed in W99, the Hall 
term has considerable effect
on the \mri growth rate ($ \sim \Omega$) in the small $\chi$ limit. This 
is because the Hall effect and collisions are intricately linked in a partially ionized plasma. Since the 
Hall effect is due to ion-neutral collision,  small wavelength
fluctuations are supressed in the Hall regime and only long wavelength 
fluctuations will grow.
We shall see that both Hall as well as Ambipolar diffusion will 
become important in the small $\chi$, high frequency limit,
suggesting that the presence of ion inertial effect destabilizes the 
weakly ionized disc at all wavelengths.
\section{Energetics of the disc}
Before we discuss numerical results, lets examine the various 
factors (viz. Lorentz force, Joule heating)
that may affect \mri. The Lorentz force, $\J\cross\B$, acts on the 
neutrals through collisions.
Making use of equation (\ref{olaw}), it can be written as
\bq
\frac{\J\cross\B}{c\,\rho\,\Omega} = \left( \chi_p\,\VB + \chi_H\,\Bh 
\cross\VB \right).
\eq
Here $\chi_P = (\sigma_P/\sigma_\perp)\,\chi$, $\chi_H =
(\sigma_H/\sigma_\perp)\,\chi\,$ and $\VB = c \Ep\cross\B/B^2$ is the
drift velocity of $\B$ through the neutrals.  The first term on the
right hand side is a measure of simultaneous acceleration and
frictional drag; viewd from a neutral frame, this force accelerates
the neutral towards $\Ep \cross \B$ velocity.  The parameter $\chi_P$
provides the strength of the collisional coupling.  With decreasing $\chi_P$, i.e. 
when the ionized medium is far from ideal MHD regime, this term may
become increasingly unimportant. Thus the modification to the ideal MHD modes will be severe in the 
small $\chi_P$ limit since collision modifies the fluid response to the magnetic tension. Note that 
in the ambipolar regime, this term is responsible for dissipation as well as feeding of energy to the neutrals.  
The second term will accelerate the neutral in the direction of $\Ep$. 

In order to understand the implications for the \mri, we need
to identify the conditions under which energy is fed to the 
fluctuations. Recall that the electric field
$\Ep = \E + \vn \cross \B/c$ is given in the neutral frame and thus,
\begin{eqnarray}
\J\cdot\E = \sigma_{\parallel} \Epa^2 + \sigma_\mathrm{P} \left( 
\Epe^2 + \vn \cdot \Epe \cross \B /c\right)
\nonumber \\
  + \frac{B}{c}\,\sigma_\mathrm{H} \left( \vn \cdot \Epe \right)  \,.
\label{ener}
\end{eqnarray}
Clearly then, the energy exchange consists of Joule heating and 
acceleration of the neutral medium. The term
$ \sigma_{\parallel} \Epa^2 + \sigma_\mathrm{P} \Epe^2 $ is the Joule heating.
This term is always positive for positive $ \sigma_\mathrm{P}$. 
However, since $\sigma_P$ may become negative near resonance and the Ohmic term $\sigma_P\,\Epe^2$
may feed rather than dissipate energy. Therefore, in the ambipolar regime, fluctuations may grow. The terms $ 
\vn \cdot \Epe \cross \B$ and $\vn \cdot \Epe$, for ambipolar and Hall respectively, corresponds to the feeding or, 
extraction of the kinetic energy by the Lorentz force. Therefore, depending upon the 
sign of the kinetic energy terms, the Lorentz force may facilitate either growth or damping
of the \mri.

\section{Results}
We shall solve the dispersion
relation (\ref{drf}) numerically in various limiting cases and discuss 
modifications due to ion inertia.
In the absence of ion inertia, various $\chi$ limits and its effect 
on the \mri have been discussed in detail in W99.
We assume that electrons are frozen in the magnetic field, i.e. $\beta_e = \infty$.
In this limit, $\sigma_P^0/\sigma_H^0 = \beta_i$. 
Therefore, we shall solve the dispersion relation (\ref{drf}) by varying 
key parameters $\chi$, $\beta_i$ and $\nu = \nu_{in}/\Omega$. 
In accretion disc environment, the value of $\beta_i$, may vary in a 
wide range. we shall choose $\beta_i$ between
$0.1$ and $1$. Although higher value of $\beta_i$ can be chosen, 
the growth rate of \mri will be very small. The parameter $\nu_{in}/\Omega$ is similarly varied in 
a wide range. 
\subsection{Variation of $\beta_i$ for $\chi = 0.1,\,\nu=1$}
In Fig. 2(a) we plot the growth rate by varying $\beta_i$ while keeping $\nu=1$ and $\chi=0.1$ fixed. 
With the decreasing $\beta_i$, the parameter window
of \mri extends towards short wavelength  and the growth rate exceeds 
ideal MHD limit for $\beta_i = 0.1$. With the decrease in $\beta_i$ when $\omega_{ci} <
\nu_{in} \sim \Omega$, the mode grows upto
$0.92\,\Omega$ for $\beta_i = 0.1$. This can be possibly attributed to the fact that owing to the 
faster collisional (and Keplerian) time scales in comparison with the gyration time, rotational free
energy becomes available to the fluctuations at the collisional time scale.
For further decrease in the value of $\beta_i$ to $0.01$, the maximum growth rate  decreases. 
This behaviour indicates that if $\beta_i$ is increased beyond some critical value, the 
Hall diffusion is dominated by the ambipolar diffusion. The growth rate
for $\beta_i = 1$ curve is small. This regime correspond
to $\omega_{ci} \sim \nu_{in} \sim \Omega$, i.e. the rate of ion 
gyration and collision with the neutrals is comparable with  the
rotational frequency. With the increase of $\beta_i$, the growth rate 
decreases and disappears altogether for very
large $\beta_i$. Thus, Hall effect (caused by the ion inertial and collisional 
effects), starts dominating the ambipolar diffusion and the mode
starts growing. Further decrease in $\beta_i$ and increase in Hall 
conductivity reduces the growth rate to $0.82 \Omega$. The
collisional effect weakens with decreased $\beta_i$ and thus, the parameter 
window operates in both small and large wavelength
regimes.
\subsection{Variation of $\chi$ and $\nu$ for fixed $\beta_i$}
\begin{figure}
       \includegraphics[scale=0.45]{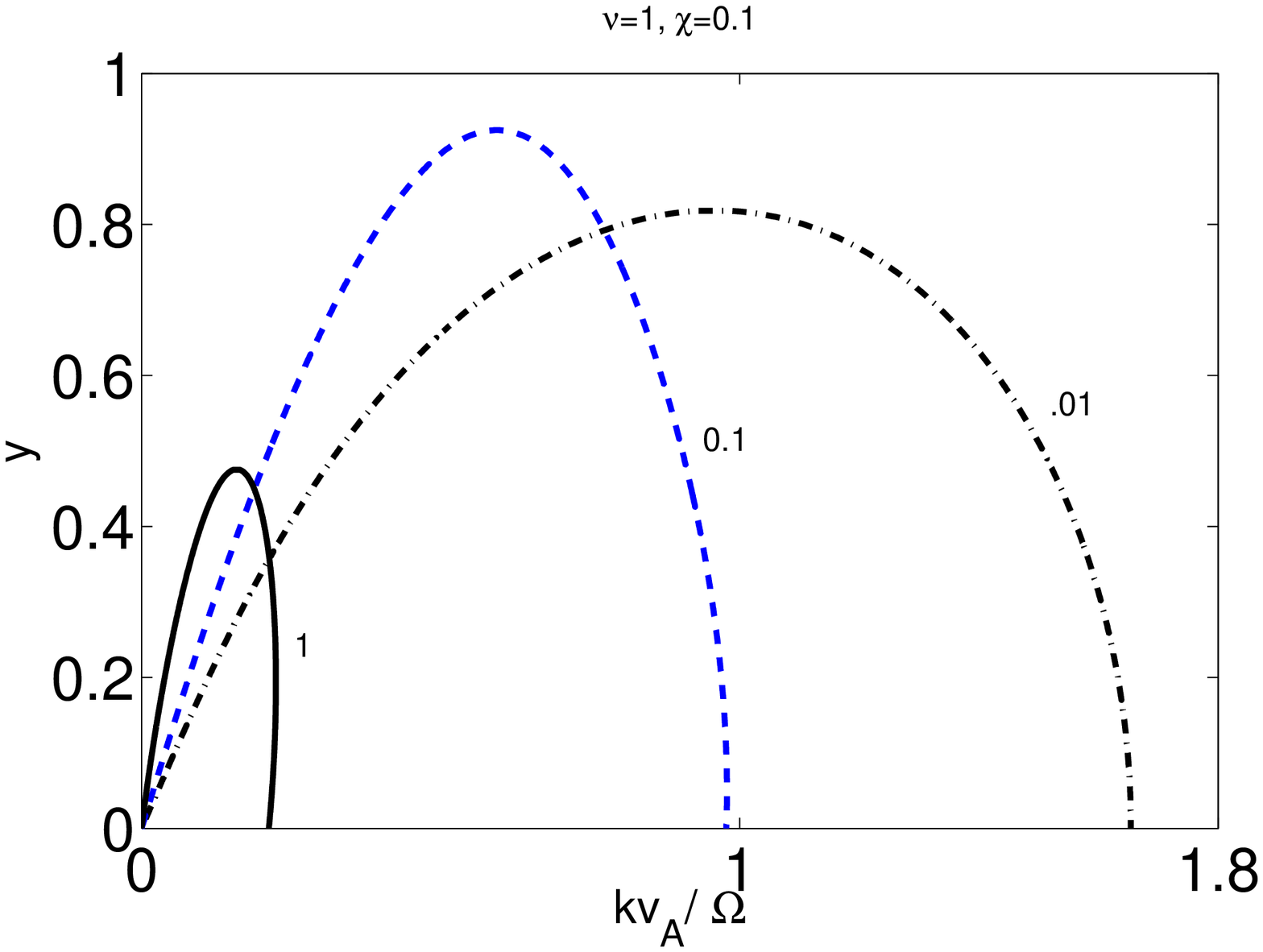}
       \includegraphics[scale=0.45]{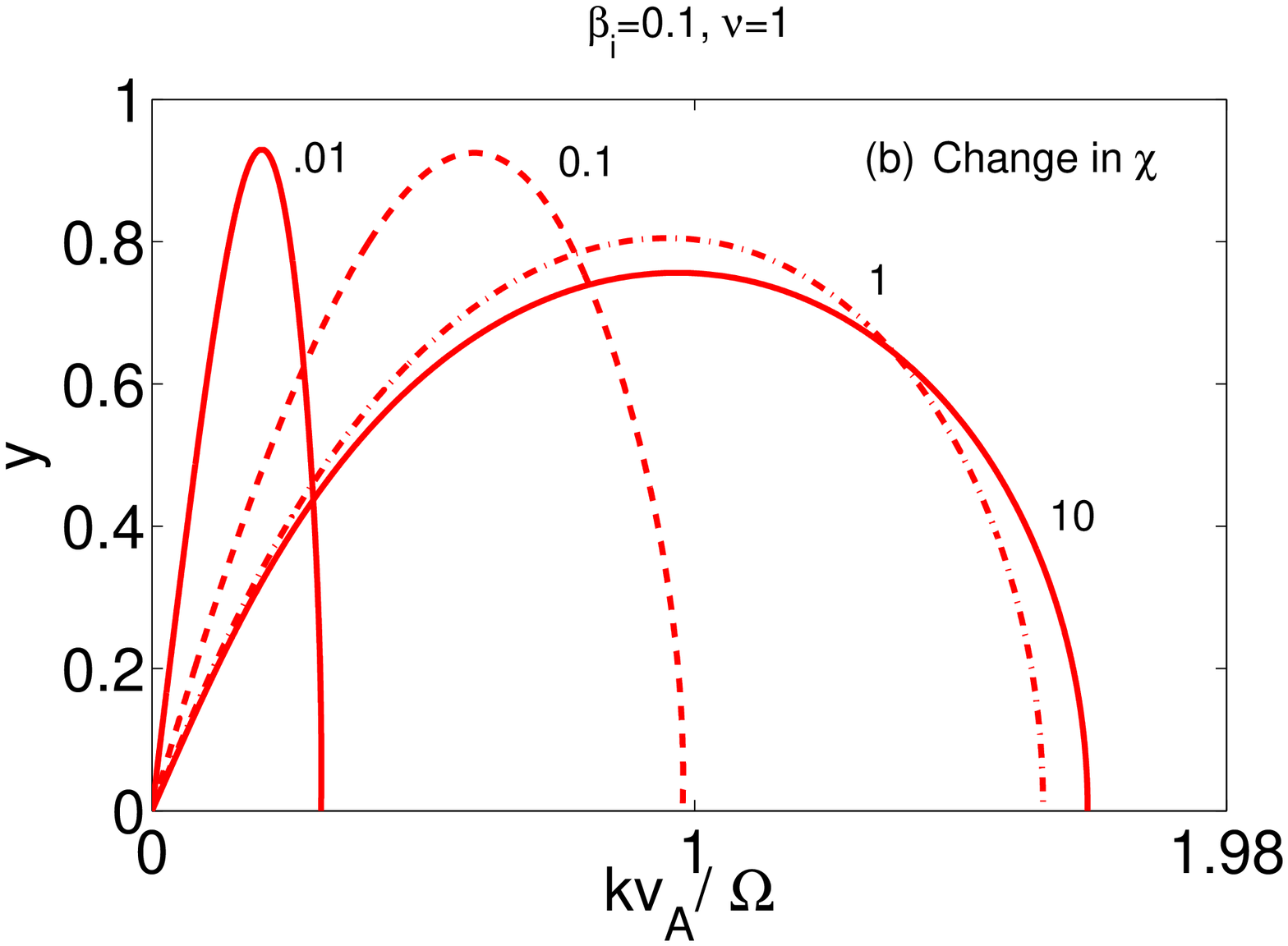}
       \includegraphics[scale=0.45]{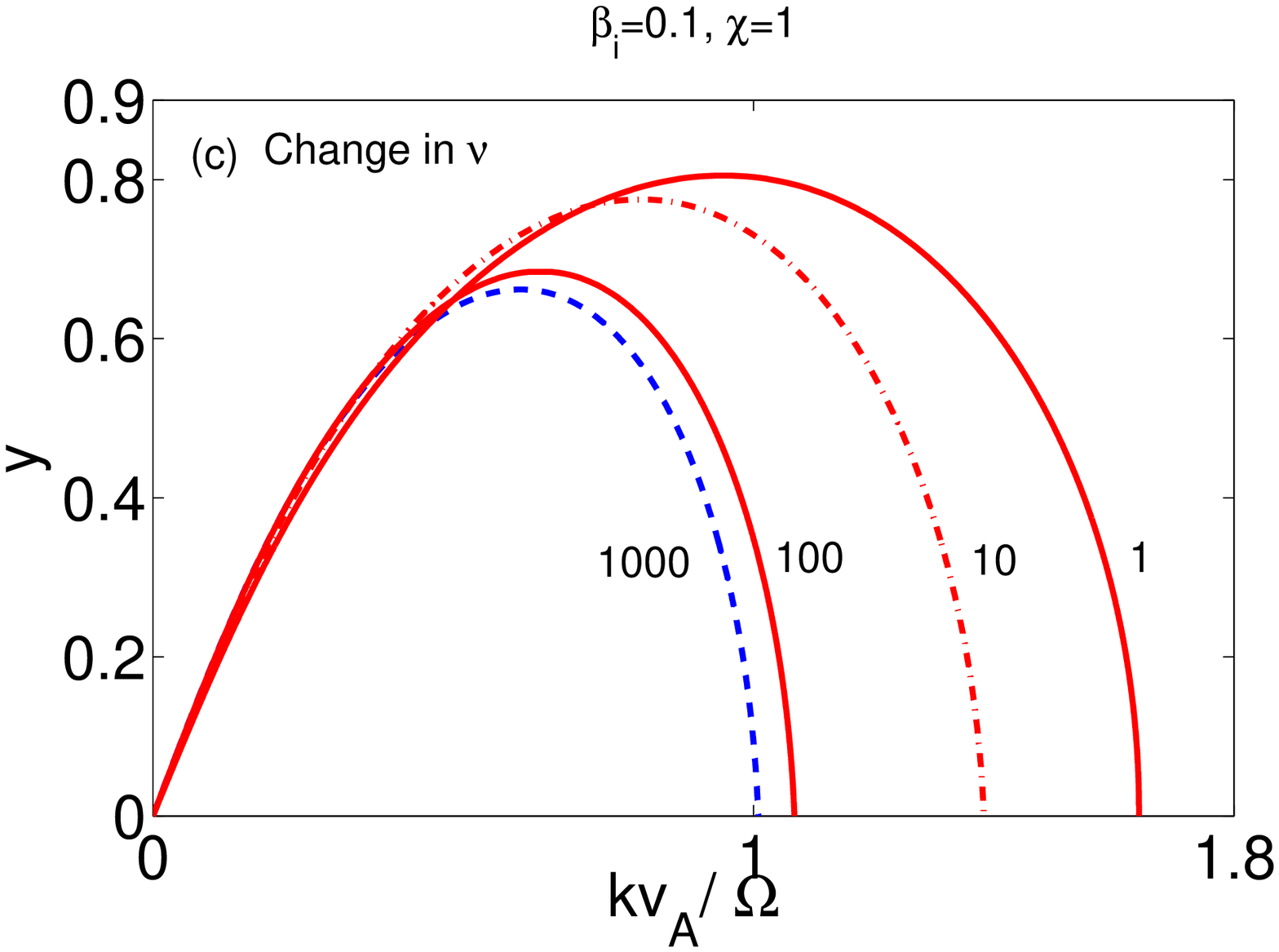}
      \caption{\large{The \mri growth rate for $\sigma_P =1$ for varying 
$\beta_i$ and $\sigma_H$ with
fixed $\nu = 1$  and $\chi = 0.1$ is plotted against the wavenumber. 
The number against the curve is the value of ion Hall $\beta_i$ for 
$\sigma_H = 1,\, 10$ and $100$. In figure 2(b) we hold $\nu$ 
fixed and give growth rate for various $\nu$. In
Fig.2(c) we hold $\chi$ fixed and vary $\nu$.}
}
\end{figure}

In Fig. 2(b) the growth rate is given for varying $\chi$ and fixed 
$\beta_i = 0.1,\,\nu = 1$. With the decreasing $\chi$ (i.e. increasing collisional coupling), the wavenumber 
window of the instability shifts towards long wavelength,
consistent with the fact that non-ideal
MHD effects start playing an increasingly important role for smaller 
$\chi$. The result is similar to W99. This result is also
in agreement with \cite{bb}. We see that with the decreasing $\chi$ 
the growth rate remains unchanged. Only change is in
the wavenumber window that shifts towards the long wavelength 
consistent with the Hall dominated result of W99.

In fig. 2(c) we plot the growth rate for $\chi =1$ and $\beta_i = 
0.1$ for varying $\nu$. The results are along the expected
line. With the increase in collision, the growth rate decreases due 
to increased role of dissipation and the parameter
window of instability shifts towards long wavelength. When the 
ion-neutral collision rate is comparable (or smaller) to the
rotational frequency, the MRI is unaffected by the collision. This is 
because the rate of dissipative loss of the energy
is comparable or slower than the rotational time scale $\nu_{in}^{-1} 
\sim \Omega^{-1}$. Thus dissipation stops affecting the growth
rate and it saturates around $\sim 0.75$. Further decrease of $\nu$ 
does not affect the growth rate.
\subsection{General limit with $\beta_i =  1$}
\begin{figure}
      \includegraphics[scale=0.45]{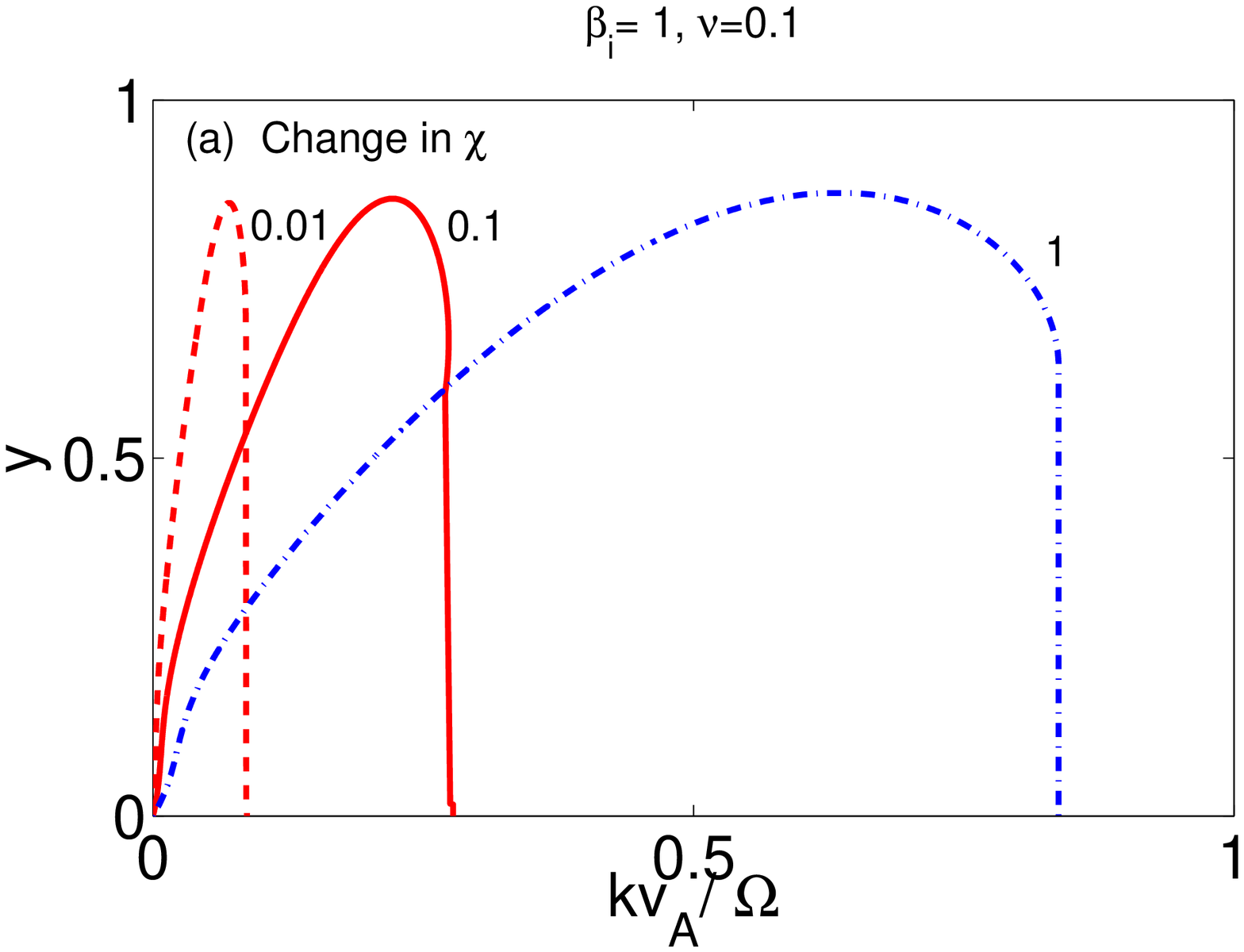}
      \includegraphics[scale=0.45]{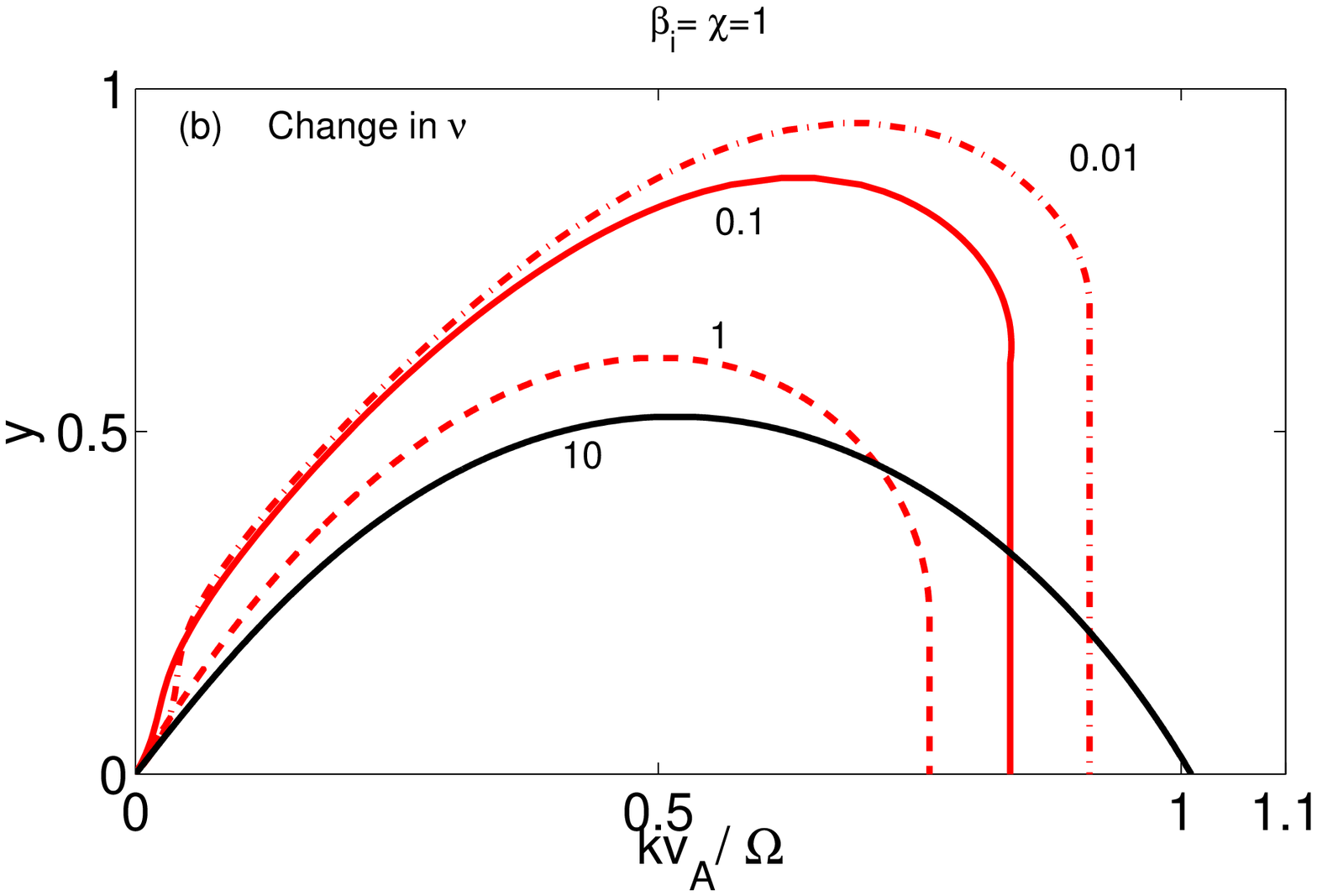}
      \caption{\large{As for Fig.2 but for $\sigma_P = \sigma_{H} = 1$}}
\end{figure}
In figure 3(a) we plot the growth rate for the positive orientation of the magnetic field with respect to the 
rotation axis, i.e. $s =1$.  The results are similar 
to the known results of W99. However, there are some
interesting differences towards small $\chi$ limit. Whereas, the 
wavenumber window of \mri shifts towards longer wavelengths
in small $\chi$ limit, the growth rate of instability is not very 
sensitive. The ion inertia is able to
provide the free energy to the fluctuations that can counterbalance 
the dissipative losses in small $\chi$ regime where
collision coupling between ion and neutral is very strong.  Thus, 
with the decrease in the value of $\chi$, small wavelength
fluctuations are all suppressed leaving large wavelength modes to 
grow at $\sim 0.82\,\Omega$.

In figure 3(b) we plot the growth rate against $\nu_{in}/\Omega$. 
We see that when $\nu_{in} \ge \Omega$ i.e., when
collision time $t_c \equiv \nu_{in}^{-1}$ is smaller than the 
rotational time $t_r \equiv \Omega^{-1}$, and the free energy
is dissipated by the collision. The growth rate of fluctuations decreases.
We see from the plot that with increasing $\nu$ the \mri growth rate 
reduces significantly. However, the growth rate becomes
insensitive beyond $\nu=10$. This indicates that in the large $\nu$ 
limit, ion inertial effect, namely Hall effect starts
cancelling dissipation and thus, growth rate becomes insensitive to 
any further increase of $\nu$. In the opposite
limit, i.e. when $\nu \le 1$ (or, $t_r < t_c$), the energy available 
to the fluctuation is at the rotational time scale and, infrequent, 
slow ion-neutral collision is unable to influence the growth of the 
instability. At $\nu = 0.01$, the growth rate becomes
maximum $\sim 0.9 \Omega$ and any further decrease in $\nu$ do not 
change the growth rate significantly.
\subsection{Weak ambipolar limit with $\beta_i = 0.1,\,s = -1$}

In figure 4(a) and 4(b) we plot the growth rate for $B_Z < 0$  by 
varying $\chi$ and $\nu$ respectively. In figure 4(a) the growth rate
is slightly smaller than for corresponding case in Fig. 3(a) for s=1. 
For $\chi\le 0.1$ the
growth rate is insensitive to the changing value of $\chi$ implying 
that if collision frequency is an order
of magnitude smaller than the Keplerian frequency, the ambipolar 
effects are entirely compensated by the Hall
and the growth rate remains constant. The  curves are similar to W99 except that 
the mode exists for much
smaller value of $\chi$ than was the case in W99. Also, the growth 
rate is larger. The growth rate increases
with increasing $\chi$ and attains maximum value for $\chi =10$. If 
$\chi$ is increased further, there is no
change in the growth rate. The rate at
which rotational energy becomes available and dissipation operates 
become comparable. For $\chi>1$, the ideal
MHD limit is approached and thus the effect of collision diminishes. 
Thus for $\chi = \infty$ the growth rate
is maximum.

In figure 4(b) the growth rate is not very sensitive to change in 
$\nu$ except when it
becomes large. For large $\nu$ the growth rate reduces in comparison 
with the small $\nu$ values and parameter
window extends towards smaller wavelength.
\begin{figure}
      \includegraphics[scale=0.45]{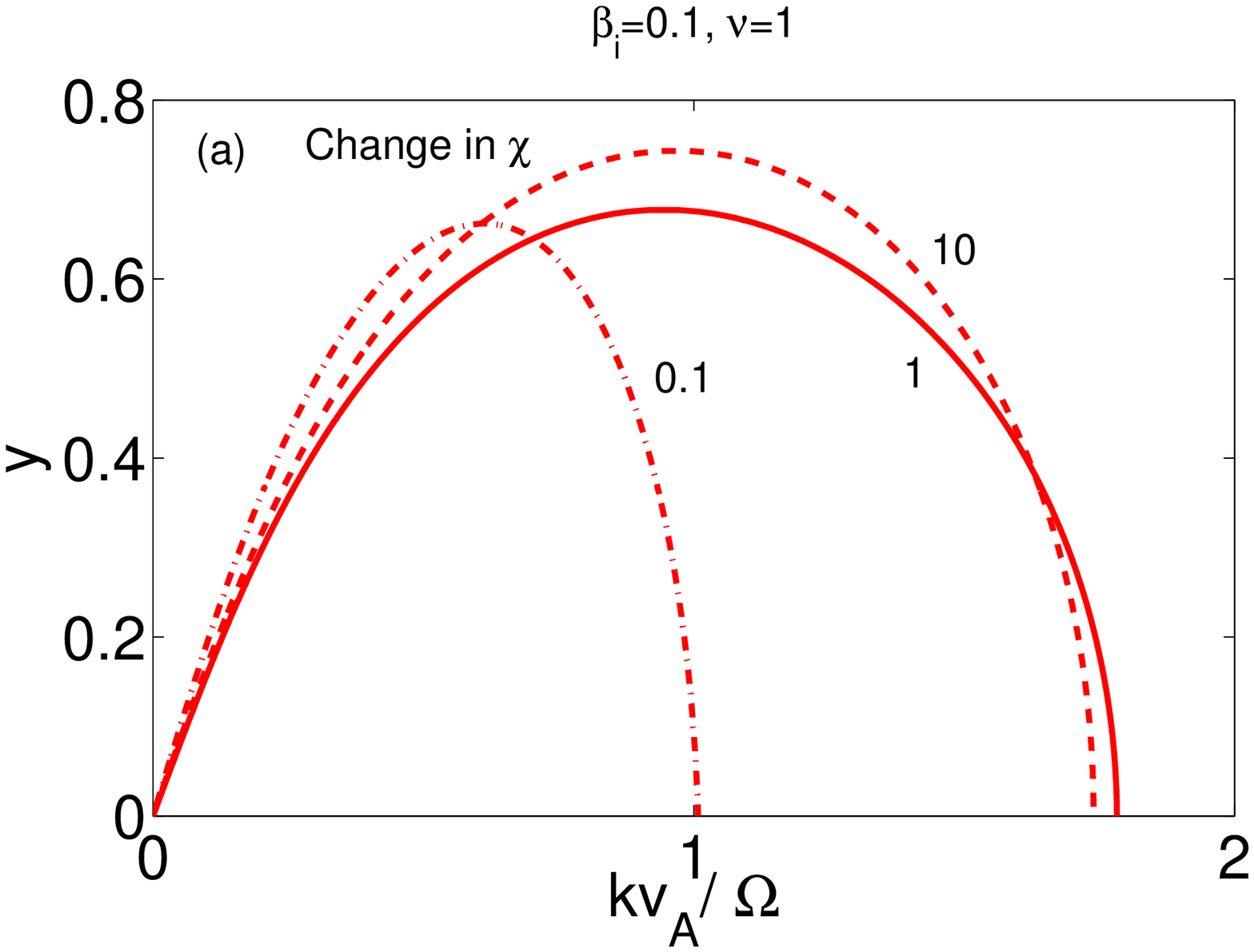}
      \includegraphics[scale=0.45]{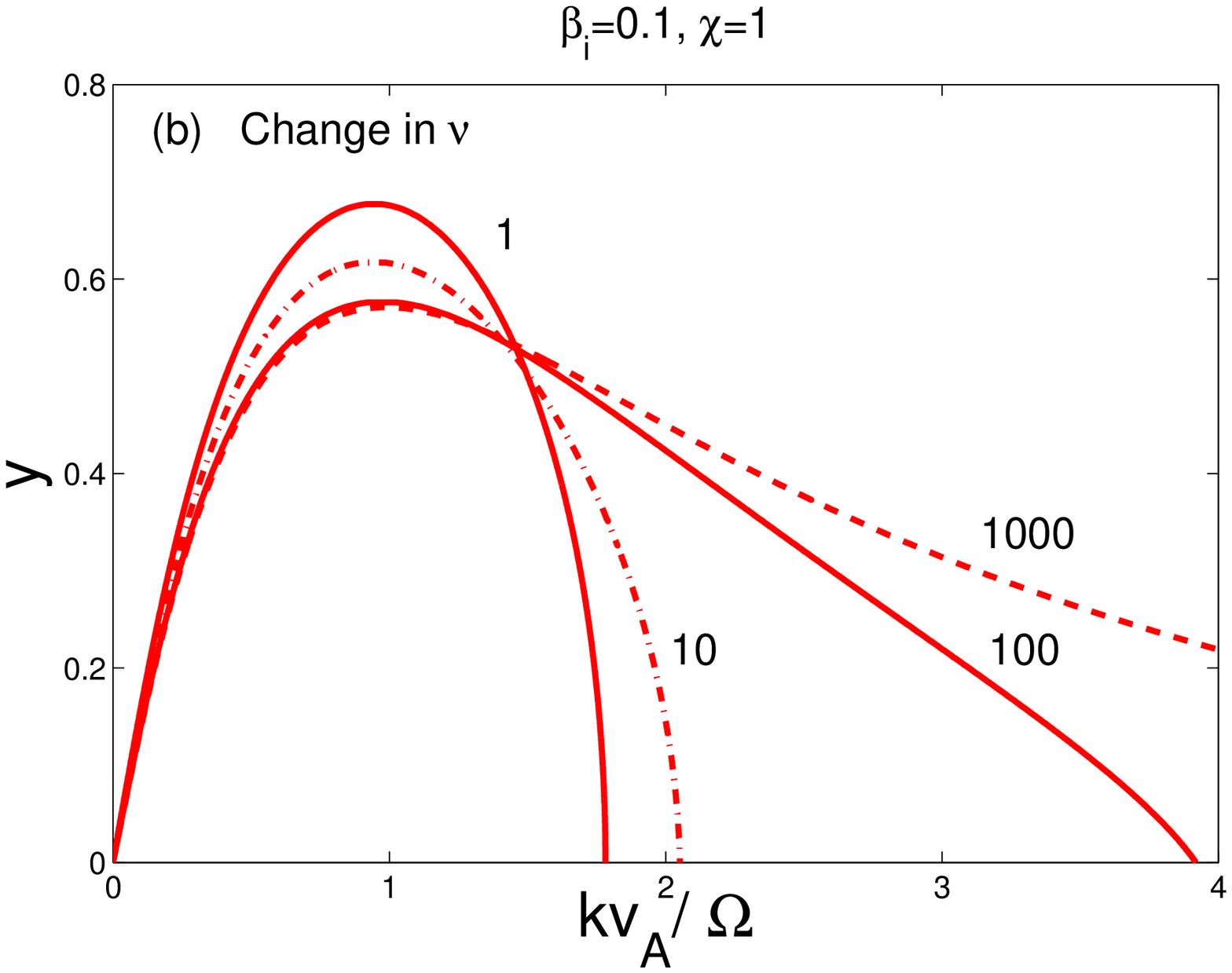}
      \caption{\large{As for Fig. 2(b), 2(c) but for $s=-1$.}}
\end{figure}
\subsection{General limit for $s = -1$ and $\beta_i = 1$}
In Fig. 5(a) we plot the growth rate for $s = - 1$ for varying $\chi$. The
growth rate is insensitive when $\chi < 1$. Small $\chi$ is a 
measure of departure from ideal MHD and we note
that Hall effects play important role for $\chi < 1$. Since for $s = 
-1$ the sign of the wave helicity $\bmath{\Omega}\cdot\dB$ is negative since \alf wave is propagating in the negative 
direction (\ref{hel2}). Thus, the
increase in the non-ideal effect, manifested through Hall terms, does 
not have any bearing on the growth rate. With
the increase of $\chi$ the \mri growth rate approaches ideal limit.

In Fig. 5(b) the variation of growth rate with $\nu$ is given. For 
$\nu \le 1$ there is no change
in the growth rate. For $\nu > 1 $ the instability is damped due to 
dissipation. The sign of the helicity
ensures that non ideal effect do not feed the energy to the 
fluctuations. Thus we see the decrease in the growth rate
with increasing $\nu$.
\begin{figure}
      \includegraphics[scale=0.45]{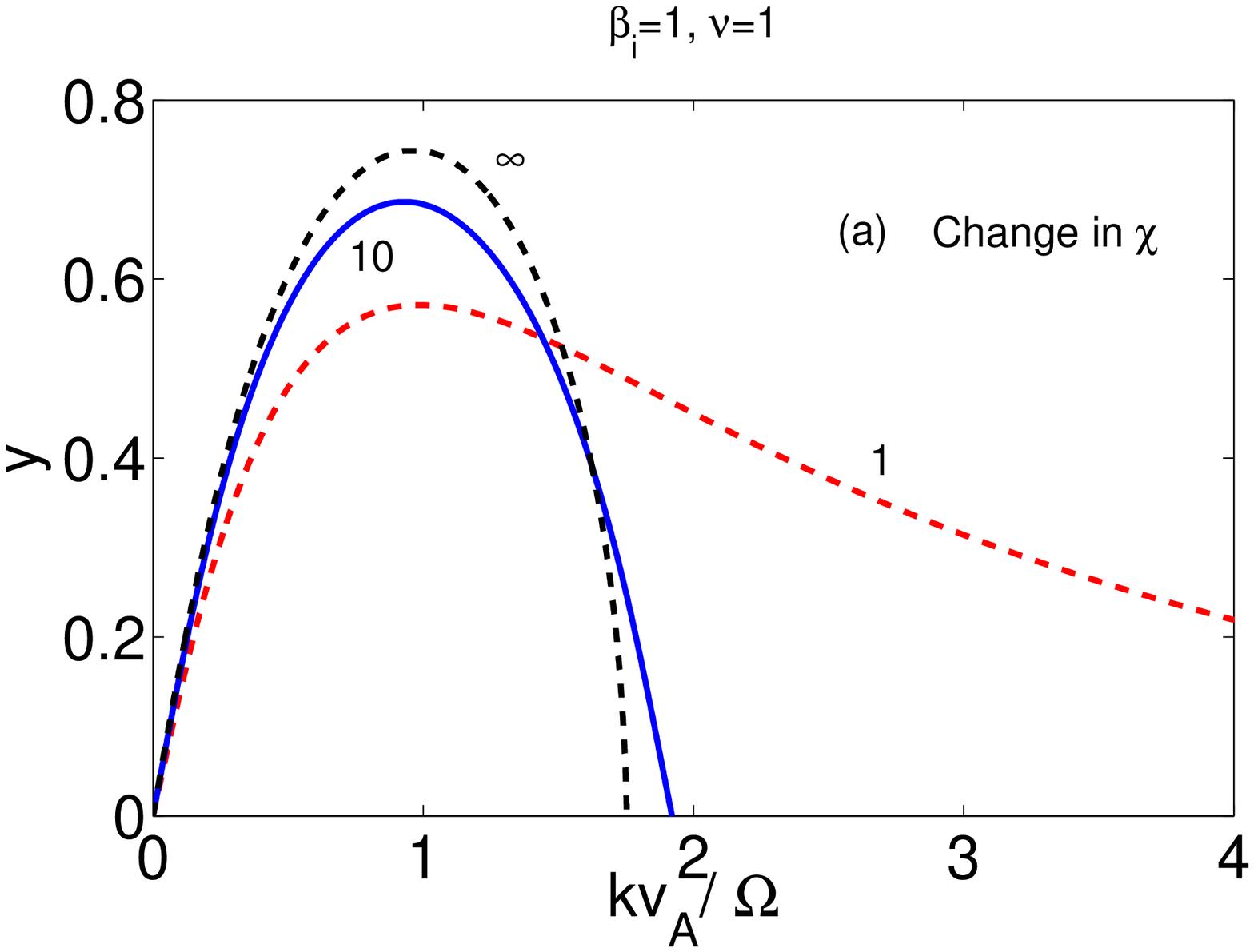}
      \includegraphics[scale=0.45]{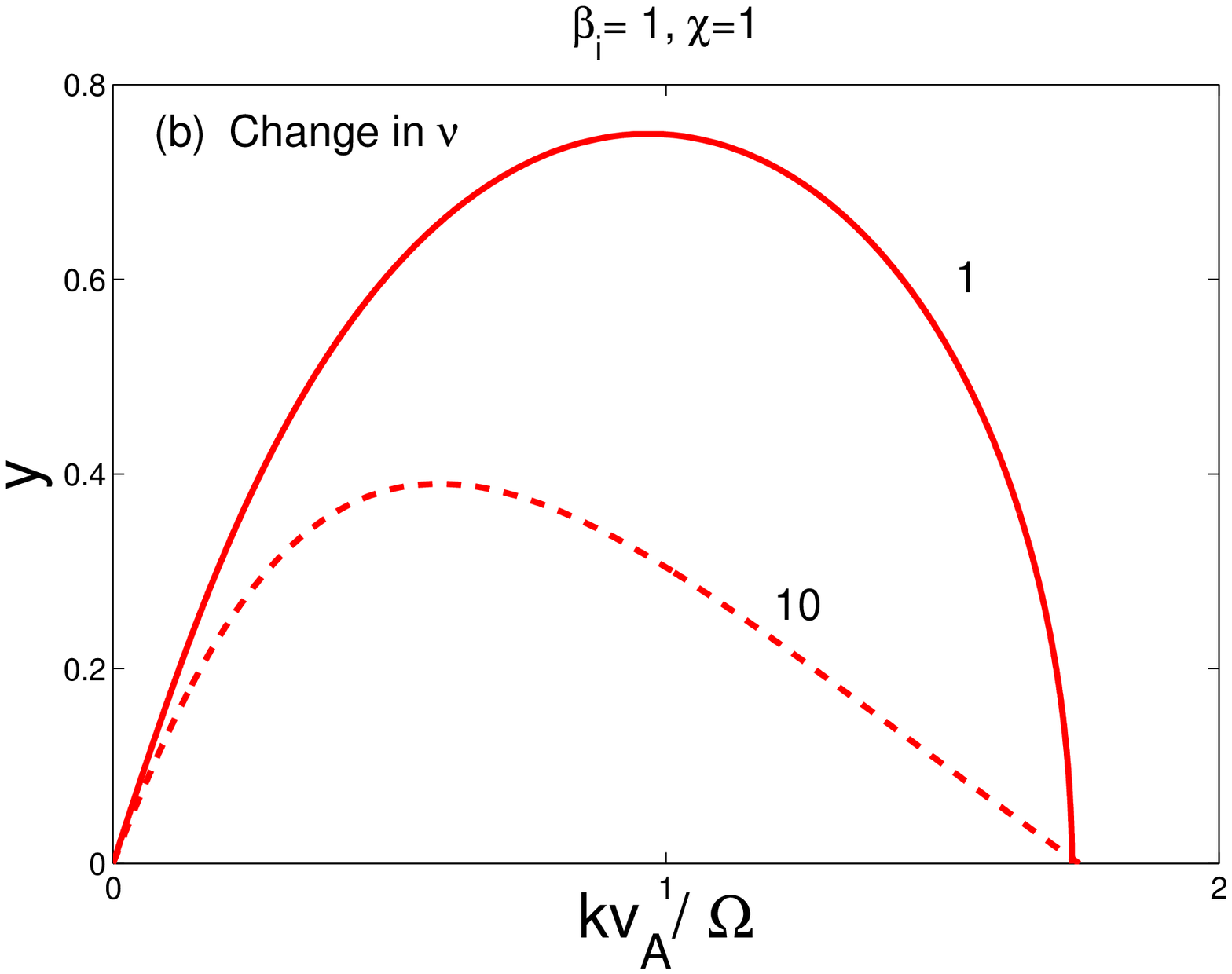}
      \caption{\large{As for Fig.2 but with $s = -1$.}}
\end{figure}
Therefore, we see that in the general case when the ambipolar 
conductivity may change, the growth rate of the instability becomes larger than the rotation frequency and the 
instability operates at long wavelengths. Thus, the
ion-inertial effect introduces entirely new feature to the dynamics 
of a weakly magnetized disc. Not only it  changes the length scale over which the 
instability can operate but also, how fast it can operate. These feature makes ion inertial effect very important for the 
application to the protostellar discs.
\section{Applications}
The modification to the \mri by ion inertia may have wide ranging application in the astrophysical discs. Before discussing the 
application of the results, we shall note that the parameters $\chi$, $\nu$ and $\beta_i$ are not independent but 
are related by the following equation 
\bq
\frac{\rho_i}{\rho_n} = \sqrt{1 + \beta_i^{-2}}\,\left(\frac{\chi}{\nu}\right),
\eq
where $\nu \equiv \nu_{in}/\Omega$ is the normalized collision frequency and $\chi$ is the measure of non-ideal MHD 
effects. Therefore, the choice of $\chi$, $\beta_i$ and $\nu$ constrains the ratio $\rho_i/\rho_n$ and hence, the level of 
fractional ionization. 

At densities relevant to cloud cores and protostellar discs (densities $ \ge 10^{11} \mbox{cm}^{-3}$), 
grains are the dominant charge carriers and their presence can significantly alter the dynamics of the disc. 
The ionization fraction is strongly affected by the abundance and size distribution of the grains through the 
recombination process on the grain surface. Near the midplane of the PPDs grains are the main charged constituent \citep{w5}. 
We shall assume a Keplerian frequency for the
minimum mass solar nebula ($0.1\,M_{\sun}$),  $\Omega \sim 
10^{-8}\,\mbox{s}^{-1}$.  Then, from equations (\ref{cf0}) and
(\ref{cg}) we write
\bq
\frac{\nu_{in}}{\Omega} \sim \frac{m_n}{m_i}\,10^{14}\,a_{-5}^2\,\left(\frac{n_n}{10^{11}\,\mbox{cm}^{-3}} 
\right).
\eq
Taking $m_n/m_i \sim 10^{-14}-10^{-15}$, we see that 
$\nu_{in}$ is comparable to the dynamical frequency and
ion-inertia becomes important. The collision of the energetic 
electron with neutrals or \mri induced turbulent
convective homogenization of the entire disc \citep{si} may allow the 
magnetic field to couple to the disc matter near the
mid plane. Therefore, the ion-inertia may modify the parameter window 
of instability near the mid-plane of the disc.

In AGNs, for example NGC 4258, a thin disc of $0.2$ pc 
diameter, bound by a central mass of
$\sim 2.1\,\times\,10^7\,\mbox{M}_{\sun}$, is rotating with a 
velocity $900\,\mbox{km}\,\mbox{s}^{-1}$ \citep{LJG}. The
observed emission emanates from an annulus of inner radius $\sim 
0.13$ pc and outer radius of $0.25$ pc. Taking $R = 0.1$ pc,
we get $\Omega \sim 10^{-10}\,\mbox{s}^{-1}$. Thus the ratio 
$\nu_{in}/\Omega \sim n_n/\mbox{cm}^{-3}$.
Taking ionization fraction $X_e = 10^{-5}$ \citep{MN} at 
$0.1\,\mbox{pc}$, we see from equation (\ref{REA}) that
$Re_A = 10^{-3}\,n_n/\mbox{cm}^{-3}$. The neutral density 
$n_n \sim 10^{7}\,\mbox{cm}^{-3}$ and thus, both ion-neutral 
$\nu_{in}$
as well as neutral-ion $\nu_{ni}$ collision frequencies are very 
large in comparision with the rotation frequency. The charged grains are negligible 
in such a disk since from $\rho_i/\rho_n = 10^{-2}$, we have $n_i/n_n \sim 10^{-14}$
for micron-sized grains. Therefore, the charged grains are abesnt in such a disk and grain 
inertia will have no effect on the disc dynamics.

Given the uncertain nature of the disc size, if we assume a disc of 
$100$ pc with a temperature gradient towards the outer edge
of the disc then the inertial effects in such a disc will be due to the charged grains 
near the core of the disc and due to the lighter ions near the surface region of 
the disc. For $\Omega \sim 10^{-8}\,\mbox{s}^{-1}$ (at $100$ pc), 
$\nu_{in}/\Omega \sim 10^{-2}\,n_n/\mbox{cm}^{-3}$, $X_e \le 1$,
we see that in the surface region of an disc, both ion and neutral 
fluid will be affected by the \mri. Thus,
ion-inertial effect may be important in exciting MHD turbulence in 
the whole disc.

Cataclysmic Variables (CV) are close binary systems with a white 
dwarf accreting material from a Roche-lobe and a companion
low mass main sequence star. The typical orbital
frequency of the CVs vary between $\Omega \sim 10^{-3} - 
10^{-5}\,\mbox{s}^{-1}$. Then  $\nu_{in}/\Omega \sim \left(10^{-5} - 
10^{-7}\right)\,\times\,n_n/\mbox{cm}^{-3}$. For $n_n \sim \left(10^5- 
10^7\right)\,\mbox{cm}^{-3}$, $\nu_{in}/\Omega  \sim O(1)$. The 
temperature
in CVs may vary in a wide range and disc can be modeled either as a 
weakly ionized plasma \citep{GM} or a completely
ionized plasma \citep{SWCR}. Clearly, both ion and neutral inertial effect 
operate on an equal footing in CVs.

The circumnuclear disc at the galactic centre has a typical constant rotation speed of $110$ Km s$^{-1}$ (\citep{GT}) 
between 2 to 4 pc. The corresponding rotational frequency at $2-4$ pc
is, $\Omega \sim 10^{-12}-10^{-13} \,\mbox{s}^{-1}$. Then the ratio between ion-neutral collison to the Keplerian roration 
frequency is
\bq
\frac{\nu_{in}}{\Omega} \approx \left( 10^{2} - 10^{3} \right)\,\times\,n_n,
\label{rcf}
\eq 
for given $\nu_{in}$ value (equation (\ref{cf2})). Hence at $2-4$ pc, the ion inertial response time, $\Omega^{-1}$ is thousand 
times slower than the collisional momentum exchange time, $\nu_{in}^{-1}$. Therefore the ability of ion inertia to modify 
the \mri at $2$ pc is unclear although $Re_A \sim 1$ for $n_i/n_n \sim 10^{-3}$. At $100$ pc where $\Omega$ drops by two orders,
$\nu_{in}$ becomes comparable to the rotational frequency and intertial effect on the \mri may become important. 
\section{Conclusions}
The paper examines the role of ion inertial effect on the 
magnetorotational instability in a weakly ionized, thin,
magnetized Keplerian disc. The vertical stratification and radial and 
azimuthal variations were neglected - an
approximation valid for the wavelengths small compared to the disc 
scaleheight. The conductivity tensor becomes
time dependent in the presence of ion inertial terms. This may result 
in the conductivity becoming negative near resonance.
Further, radial and azimuthal component of Hall conductivity will be 
different. The following results were found.

(i) The conductivities in a weakly ionized gas is in general
complex in the presence of ion inertia and may become negative near
the resonance points. The \mri gets significantly modified in the presence 
of time dependent conductivities.

(ii) In weak ambipolar regime, the presence of ion-inertial effect 
substantially modifies
the behaviour of the instability in the non-ideal ($\chi < 1$) limit. 
The maximum growth rate in the Hall dominated regime
is $\sim 0.92$ (in the units of $\Omega$) and large wavelength 
fluctuations can grow due to the inertial effect.

For a fixed $\chi$, the growth rate is maximum ($\sim 0.8$) for $\nu 
= 1$ and starts to decrease with
increasing $\nu$. Further, the parameter window shifts towards longer 
wavelength with increasing $\nu$.

(iii) When both ambipolar and hall diffusion are comparable, the 
maximum growth rate is $\sim 0.85$ and $0.95$ for
fixed $\nu$ and $\chi$ respectively. The increase in the ambipolar 
diffusion causes the fixed $\chi$ case to have
larger growth rate than fixed $\nu$ case.

To summarize, it is a common feature that large scale fluctuations 
exhibit the maximum growth rate of the \mri
when the ion inertial effects are included in the dynamics. This may 
have important implication on the onset of turbulence in weakly ionized discs. 
For example, in PPDs, when grains are dominant ions, the grain inertial effect 
will significantly modify the \mri growth rate and thus, will play an important 
role in the onset of hydromagnetic turbulence. In AGNs also, grain will play important 
role. In CVs grain will provide the ion inertia away from the surface of the dwarf novae whereas, 
lighter ionized elements will provide the inertial effect close to the surface of the disc. Therefore, 
in CV discs, inertial effect may be important all across the diak and inertia modified \mri may effect 
the whole disc. In circumnuclear discs, the effect of ion inertia may be important far away from the centre of 
the disc. All in all, ion inertia seems to play an important role on the onset of \mri.

Present work inevstigates the role of ion inertia in the presence of an axial magnetic magnetic field. It will 
be interesting to investigate the role of ion inertia on the \mri for a more general field geometry, particularly in 
the context of profile independent destabilizing feature of ambipolar diffusion \citep{kz}. We shall leave this problem 
for future consideration. 

\section*{Acknowledgments}
We thank an anonymous referee for very useful comments that improved
the presentation of the paper. This reserach was funded by the Australian Research Council (ARC).

\end{document}